\documentclass{ar-1col}
\usepackage[numbers]{natbib}
\usepackage{url}
\usepackage{bm}
\jname{Annual Review of  Particle and Nuclear Science}
\jvol{AA}
\jyear{2021}
\doi{10.1146/((please add article doi))}

\begin{document}

\markboth{Kievsky et al.}{Efimov Physics}

\title{Efimov Physics and Connections to Nuclear Physics}

\author{A. Kievsky,$^1$  L. Girlanda,$^2$ M. Gattobigio,$^3$ M. Viviani$^1$
\affil{$^1$Istituto Nazionale di Fisica Nucleare, Largo Pontecorvo 3, 56100
Pisa, Italy}
\affil{$^2$Dipartimento di Matematica e Fisica "E. De Giorgi", Università del
Salento, I-73100 Lecce, Italy}
\affil{$^3$ Universit\'e C\^ote d'Azur, CNRS, Institut  de  Physique  de  Nice,
1361 route des Lucioles, 06560 Valbonne, France }}

\begin{abstract}
 Physical systems characterized by a shallow two-body bound or virtual
 state are governed at large distances by a continuous-scale invariance,
 which is broken to a discrete one when three or more particles come into
 play. This symmetry induces a universal behavior for different systems,
 independent of the details of the underlying interaction, rooted in the
 smallness of the ratio $\ell/a_B \ll 1$, where the length $a_B$ is associated
 to the binding energy of the two-body system $E_2=\hbar^2/m a_B^2$ and $\ell$
 is the natural length given by  the interaction range. Efimov physics 
 refers to this universal behavior, which is often hidden by the on-set of
 system-specific non-universal effects. In this work we identify universal
 properties by providing an explicit link of physical systems to their
 unitary limit, in which $a_B\rightarrow\infty$, and show that nuclear systems belong
 to this class of universality.
\end{abstract}


\maketitle

\tableofcontents

\section{Introduction}

Studying a particular physical system we could wonder about the interactions that govern 
the underlying dynamics. Usually, the particular 
characteristics of those interactions are revealed by the properties of those systems. The case of residual interactions is especially interesting, as exemplified by the nuclear interaction, a residual interaction of Quantum 
Chromodynamics (QCD), or by molecular structures built under residual effects of  Quantum Electrodynamics (QED). We can imagine situations in which the residual interaction places the
system in a particular energy region where the characteristics of the interaction become unimportant (we may consider this as a fine tuning). 
Similar situations could occur when a system is subject to a suitable external field. Following these ideas, and focusing on a 
non-relativistic theory, we can design a short-range tunable potential describing a two-particle 
system with mass $m$ and refer to the unitary window as the region
in the space of the potential parameters such that the scattering length $a$ reaches a value close to 
infinity. When $a$ is large the two-body system has a shallow (real or virtual) bound state whose 
binding energy is governed by the scattering length, $E_2\approx\hbar^2/m a^2$.
Its shallow character is defined with respect to the typical energy of the system, 
$\hbar^2/m \ell^2$, where the typical length of the system $\ell$ could be for example the potential range. 
The limit $\ell /a \rightarrow 0$ can be reached in two ways: the scattering length going to infinity 
(unitary limit) or the interaction range going to zero (zero-range limit or scaling limit). 
When $\ell /a \ll 1$ the system is inside the unitary window, a particular region
in which universal behavior can be observed allowing for a common description of 
totally different systems, ranging from nuclear physics to atomic physics or down in scale to 
hadronic systems.

Weakly bound systems define a class of universality; 
the particles stay most of the time outside the interaction range and many of their properties can 
be explained in terms of the probability of being inside the classically forbidden region. The fine tuning
of the potential parameters, needed to bring a system inside the unitary
window, can be realized in laboratories using external fields, like for trapped cold atoms with Feshbach 
resonances~\cite{chin:2010_Rev.Mod.Phys.} or can be naturally produced. 
There are a few natural systems located inside this 
window, one is the dimer of two helium atoms. In fact the $^4$He$_2$ molecule has an extremely low 
binding energy, $E_2\approx 1\,$mK, several orders of magnitude smaller than the typical interaction 
energy~\cite{luo:1993_J.Chem.Phys.}, $\hbar^2/m r_{vdW}^2\approx 1.5\,$K, given in terms of its
van der Waals length $r_{vdW}=5.08\,a_0$, $a_0$ denoting, here and in the following, the Bohr radius. Nuclear physics is another example; the deuteron binding energy
is $E_2=2.22456$~MeV, much smaller than the typical nuclear energy $\hbar^2/m\ell^2 \approx 20$~MeV,
with the interaction length given in this case by the inverse of the pion mass $m_\pi$, 
$\ell\sim 1/m_\pi \approx 1.4$~fm.

Nuclear physics is the low energy realization of QCD; in this regime QCD is a strongly interacting 
quantum field theory and therefore non-perturbative approaches are necessary. Such approaches start to 
appear in the form of Lattice QCD (LQCD) \cite{beane:2006_Phys.Rev.Lett.,ishii:2007_Phys.Rev.Lett.,yamazaki:2015_Phys.Rev.D,aoki:2012_Prog.Theor.Exp.Phys.,beane:2011_Prog.Part.Nucl.Phys.}, however detailed computations of nuclear properties with 
these techniques seem at present not yet feasible. In recent years it has been realized that the interaction 
among nucleons can be constructed in an Effective Field Theory (EFT) approach exploiting the 
symmetries of QCD \cite{weinberg:1990_Phys.Lett.B,weinberg:1991_Nucl.Phys.B,ordonez:1992_Phys.Lett.B,ordonez:1994_Phys.Rev.Lett.,vankolck:1994_Phys.Rev.C,vankolck:1999_Prog.Part.Nucl.Phys.,bedaque:2002_Ann.Rev.Nucl.Part.Sci.,epelbaum:2006_Prog.Part.NuclearPhys.,epelbaum:2009_Rev.Mod.Phys.,machleidt:2011_PhysicsReports,hammer:2020_Rev.Mod.Phys.}. In the limit of zero-mass light quarks the Chiral Symmetry appears, whose 
spontaneous breakdown gives rise to  Goldstone bosons,  the $\pi$-mesons. The mass of the pion $m_\pi$ is different from zero because of the soft explicit breaking term introduced by the  masses of the up and 
down quarks, but  is still much lower than the typical hadronic masses. Another interesting limit in QCD can be reached realizing that the mass of
the pion is close to a critical value at which the nucleon-nucleon scattering
lengths diverge~\cite{braaten:2003_Phys.Rev.Lett.,konig:2017_Phys.Rev.Lett.}. The $^1S_0$ (singlet) $a_s$ and $^3S_1$ 
(triplet) $a_t$ scattering lengths are functions of the up and down quark masses, or equivalently 
of $m_\pi$ which is related to the quark masses by the
Gell-Mann-Oakes-Renner relation~\cite{gell-mann:1968_Phys.Rev.}. It has been shown that for $m_\pi\approx 200$~MeV both scattering 
lengths diverge ~\cite{epelbaum:2006_Eur.Phys.J.C,beane:2002_Nucl.Phys.A}. At the
physical point, $m_\pi\approx138$~MeV, the values of the two scattering lengths are 
$a_s \approx -23.7$~fm and $a_t\approx 5.4$~fm, still appreciably larger than
the typical interaction length $\ell\approx 1.4$~fm.

A model-independent description of the physics inside the unitary window  is
given by an EFT based on the clear separation of scales between the typical
momenta~$Q \sim 1/a$ of the system and the underlying high momentum scale $\sim 1/\ell$
\cite{hammer:2020_Rev.Mod.Phys.,vankolck:1999_Nucl.Phys.A,bedaque:1999_Phys.Rev.Lett.,bedaque:1999_Nucl.Phys.A}.
This condition is well fulfilled in both systems, atomic helium and  nuclear physics. In the latter case this approach is known as pionless-EFT
\cite{vankolck:1999_Nucl.Phys.A,kaplan:1998_Phys.Lett.B,birse:1999_Phys.Lett.B,chen:1999_Nucl.Phys.A}.
Using such an EFT, if the power-counting is correct~\cite{vankolck:1999_Nucl.Phys.A,kaplan:1998_Phys.Lett.B,epelbaum:2017_Nucl.Phys.B,epelbaum:2018_Commun.Theor.Phys.,griesshammer:2005_Nucl.Phys.A}, one can 
systematically improve the
prediction of the observables. For instance, at low energies with $E=\hbar^2 k^2/m$, 
the $s$-wave phase shift $\delta$ determined by the effective range expansion (ERE) 
\cite{bethe:1949_Phys.Rev.} 
\begin{equation}
    k\cot\delta = -\frac{1}{a} + \frac{1}{2} r_e k^2  \,,
    \label{eq:ERE}
\end{equation}
can be reproduced by such an expansion~\cite{vankolck:1999_Nucl.Phys.A}. The leading order (LO) 
term captures the information encoded in the scattering length $a$, whereas the finite-range nature of the interaction, represented by the effective range $r_e$, constitutes the next-to-the-leading 
order term (NLO). Inside the unitary window, there is an energy pole
close to the two-particle threshold relating the scattering and bound state properties.
The extension of the ERE to the negative energy pole results in 
\begin{equation}
     \frac{1}{a_B}  = \frac{1}{a} +\frac{1}{2}\frac{r_e}{a_B^2},  \,
    \label{eq:ERE2}
\end{equation}
where $E_2=\hbar^2/m a_B^2$ defines the energy length $a_B$. It could be positive (bound state) or 
negative (virtual state). Moreover, $|a_B| \gg r_e \sim \ell $ and 
$a\sim a_B$, so the ratio $r_e/a_B$ represents a small parameter. 
In this energy region the two-body system is dominated by a continuous scale
invariance (CSI) governed by the control parameter $a_B$ with violations of the
order of $r_e/a_B$. 

The most remarkable property of systems at the unitary limit shows up at 
three-body level through the Efimov
effect~\cite{efimov:1970_Phys.Lett.B,efimov:1971_Sov.J.Nucl.Phys.}.
The CSI in the two-body system is dynamically broken at the level of three bodies  
into a discrete scale invariance (DSI).  When the
strength of the two-body interaction is such that there is a bound state at
zero-energy, an infinite tower of geometrically distributed energy states appears
in the three-body system with the energy threshold $E_3=0$ as an accumulating
point. The energy ratio of successive levels 
$E_3^{n+1}/E_3^{n}=e^{-2\pi/s_0}$ is a universal constant,
 with $s_0$ depending on the mass ratio of the constituents; for three equal bosons
$s_0\simeq1.00624$ so that $e^{-2\pi/s_0}\simeq (1/22.7)^2$.
The anomalous breaking of the symmetry gives rise to an emergent scale at the
  three-body level which is usually referred to as the three-body parameter
  $\kappa_*$, giving the binding energy $\hbar^2\kappa_*^2/m$ of a reference state
  belonging to the tower of states at the unitary point.

This effect, predicted by V. Efimov around 50 years ago,
was observed 35 years after its prediction by the group of R. Grimm~\cite{kraemer:2006_Nature}. 
An enormous amount of work, experimental as well as theoretical, has been, and still is, dedicated to
study this phenomenon. An introduction to this sector of research can be found
in the following reviews and references therein
\cite{braaten:2006_PhysicsReports,hammer:2007_Eur.Phys.J.A,platter:2009_Few-BodySyst,
ferlaino:2010_Physics,ferlaino:2011_Few-BodySyst.,frederico:2011_Few-BodySyst.,greene:2017_Rev.Mod.Phys.,naidon:2017_Rep.Prog.Phys.}. 
The physics associated with the Efimov
effect is called Efimov physics and the energy region in which the consequences
of this effect can be observed is called universal window, unitary window or Efimov window. Observation of universal behavior of systems belonging to this window
allows to understand better the universal dynamics as has been shown recently in
the analysis of the three- and four-neutron systems. The evidence of a low-energy tetraneutron, observed in two experiments \cite{marques:2002_Phys.Rev.C,kisamori:2016_Phys.Rev.Lett.}, has been attributed to the universal long-range tail of that system \cite{deltuva:2018_Phys.Lett.B,higgins:2020_Phys.Rev.Lett.}.
Furthermore, arguments based on the separation of scales have been recently exploited to describe halo nuclei, a sector of physics in which universal properties are expected to be observed, see Refs.~\cite{yamashita:2008_Phys.Lett.B,frederico:2012_ProgressinParticleandNuclearPhysics,jensen:2004_Rev.Mod.Phys.,hammer:2017_J.Phys.G:Nuc.Part.Phys.} and references therein. 

In order to study universal behaviour in few-boson and few-fermion systems,
the Schr\"odinger equation has been solved using two different
variational methods. For system with three and four particles we have used
the Hyperspherical Harmonic (HH)~\cite{kievsky:1997_Few-BodySyst,kievsky:2008_J.Phys.G} 
method and its unsymmetrized
version~\cite{gattobigio:2009_Phys.Rev.A,gattobigio:2009_Few-BodySyst.}.
For heavier systems we have implemented a version of the stochastic variational method 
(SVM)~\cite{varga:1995_Phys.Rev.C} using correlated-Gaussian functions as basis set. 

\section{Universal characterization of two-body systems}

The dynamics of two-body systems inside the universal window are highly
independent of the details of their mutual interaction. The systems satisfy an
approximate CSI, exactly verified in the case of a zero-range interaction. Though
the zero-range case was used many times as a first approximation to describe
systems inside the universal window, we proceed differently starting our
description from the effective range expansion, Eq.(\ref{eq:ERE2}), relating the three 
parameters that determine the low-energy dynamics of the system. 
It can be  cast in the following compact form
\begin{equation}
     r_e{a}  = 2{r_B}{a_B}  \,
    \label{eq:ERE3}
\end{equation}
where we have introduced the length $r_B=a-a_B$ which, together with the energy length $a_B$, completely determines the $S$-matrix of systems having one bound state~\cite{bargmann:1949_Rev.Mod.Phys.,babenko:2008_Phys.At.Nucl.}. In the case of a zero-range
interaction, $a=a_B$ and $r_B=0$. To study the dynamics of the systems
inside the universal window, we make use of a two-parameter
short-range potential and consider this potential as a minimal 
low-energy representation of the two-particle interaction fixed by two low-energy data, $a_B$ (or $a$) and $r_B$ (or $r_e$).

\subsection{The characteristic potential}

\begin{figure}
 \includegraphics[scale=0.5] {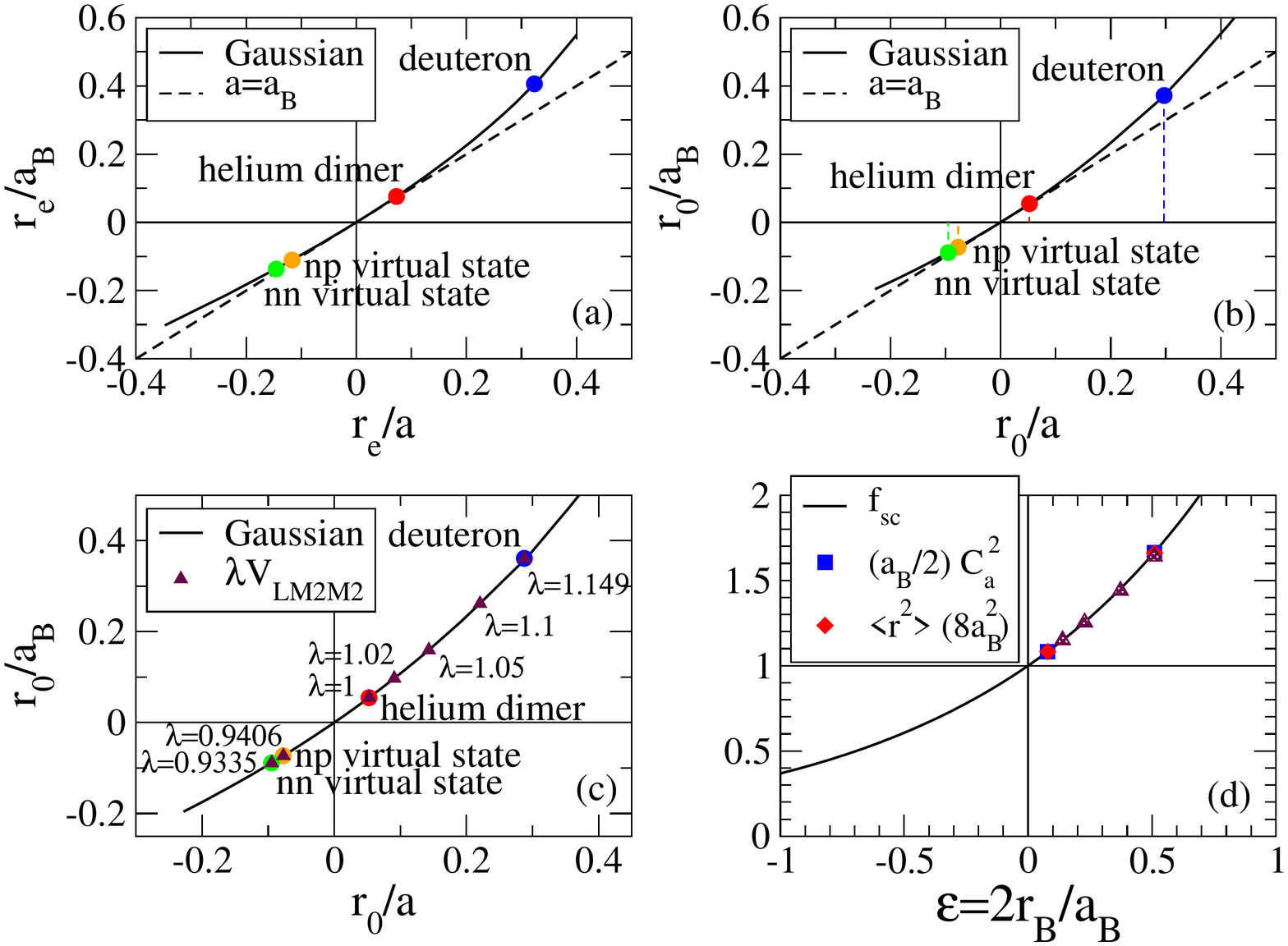}
    \caption{The inverse of the energy length as a function of the inverse of
the scattering length, both in units of $r_e$ (panel (a)) and $r_0$ (panel (b)), for a Gaussian potential. The position of selected real systems are indicated by
the solid circles. Panel (c):
position of the helium dimer (red circle), modified
helium dimers (brown triangles) and the two-nucleon systems (blue, orange and
green circles) on the Gaussian curve. Panel (d): collapse of the observables
on the scaling function.The modified helium dimers are shown as triangles.}
    \label{fig:fig1}
\end{figure}

In the following, to characterize the universal window we make use of a Gaussian potential:
\begin{equation}
  V(r) = V_0 e^{-r^2/r_0^2}\,,
  \label{eq:tbg}
\end{equation}
where $r$ is the interparticle distance, while  the strength $V_0$ and range $r_0$ are parameters useful to explore the low-energy dynamics associated with the existence of one 
(bound or virtual) state close to threshold. For bound states,
the wave function is obtained by solving the $s$-wave Schr\"odinger equation 
\begin{equation}
 \left(\frac{\partial^2}{\partial z^2} - \frac{mr_0^2 V_0}{\hbar^2}e^{-z^2} 
      -\frac{r_0^2}{a_B^2}\right) \phi_B(z) =0
\label{eq:tbsch}
\end{equation}
where $z=r/r_0$ and $\phi_B(z)$ is the reduced wave function. At zero energy,
$r_0/a_B=0$, and at large separation values $\phi_0(z\rightarrow\infty)\rightarrow
1-zr_0/a$, from which the scattering length $a$ is extracted. The zero-energy wave
function, $\phi_0$, also determines the effective range $r_e$. The small value of the ratio 
$r_e/a_B$ can be used to characterize the unitary window and, limiting the discussion to the
case of one bound state, the energy values of a generic Gaussian potential
inside the window can be organized in the single curves shown in 
Fig.~\ref{fig:fig1}, panels (a) and (b). In panel (a) $r_e/a_B$ is given as a function of $r_e/a$. 
Real systems can be placed on the figure using the corresponding values of $a$, $a_B$ and $r_e$. 
We analyze the dimer of helium atoms and the two-nucleon system. In the case of the dimer, experimental data are 
not available for all those quantities, 
so we use values obtained with one of the most widely used helium-helium interactions, 
the LM2M2 potential~\cite{aziz:1991_J.Chem.Phys.}. For the purpose of the present discussion,
results obtained with this potential are considered equivalent to experimental data. For the two-nucleon system we use the experimental values or, equivalently, the results of a realistic interaction, the AV18 potential \cite{Wiringa:1995_Phys.Rev.C}, to determine the $np$ and $nn$ low energy parameters in states $J^\pi=1^+$ and $0^+$. Using the values given in Table~\ref{tab:table1}, the four cases shown in the figure by solid circles are on top of the Gaussian curve. 

In panel (b) of Fig.~\ref{fig:fig1} the plot is reformulated in terms of the Gaussian range $r_0$,
in such a way that real systems are mapped on the Gaussian curve
through their ratio $a/a_B$. Their positions on the curve identify
the characteristic range $r_0$ indicated in the figure by the dashed lines. 
With this range, and the proper strength, a Gaussian potential reproduces simultaneously $a$ and $a_B$. 
The characteristic ranges for the deuteron, helium dimer, $np$ and $nn$ virtual
sates are given in Table~\ref{tab:table1}.

\begin{table}[]
    \centering
    \begin{tabular}{c|ccccccc}
       \hline
    He dimer & $E_2\,$[mK] &$a_B\,[a_0]$ & $C_a\,[a_0^{-1/2}]$ &
    $\sqrt{\langle r^2\rangle }\,[a_0]$  & $a\,[a_0]$ & $r_e\,[a_0]$  & $r_0\,[a_0]$ \\
         LM2M2 & -1.3035  & 182.221  & 0.108985 & 67.015  &  189.415 & 13.845 &   10.03\\
 $\lambda=1.02$& -4.0905  & 102.864  & 0.149498 & 38.979  &  110.022 & 13.396 &  9.99\\
 $\lambda=1.05$& -11.137  & 62.3388  & 0.200802 & 24.678  &  69.4483 & 12.792 &  9.94\\
 $\lambda=1.10$& -30.358  & 37.7585  & 0.277525 & 16.024  &  44.7923 & 11.937 &  9.88 \\
 $\lambda=1.149$& -57.981  & 27.3217  & 0.349857 & 12.362  &  34.2868 & 11.248&  9.86 \\
     \hline
    $NN$ & $E_2\,$[MeV] &$a_B\,$[fm] & $C_a\,$[fm$^{-1/2}$] &
    $\sqrt{\langle r^2\rangle }\,$[fm]  & $a\,$[fm] & $r_e\,$[fm] &  $r_0\,$[fm]  \\
    $np(1^+)$ & -2.2245  & 4.318  & 0.885 & 1.967  & 5.419 & 1.753 &   1.559 \\
    $np(0^+)$ & -0.066  & -25.05 &    - & -    & -23.74 & 2.77 &  1.83 \\
    $nn(0^+)$ & -0.102   & -20.19 & -    & -   & -18.90 & 2.75 &  1.795 \\
     \hline
    He dimer & $E_2\,$[mK] &$a_B\,[a_0]$ & $C_a\,[a_0^{-1/2}]$ &
    $\sqrt{\langle r^2\rangle }\,[a_0]$  & $a\,[a_0]$ & $r_e\,[a_0]$ & $r_0\,[a_0]$  \\
    $\lambda=0.9406$ &  -    & -139.631 & -    & -   & -132.318 & 15.445 & 10.19 \\
    $\lambda=0.9335$ &  -    & -115.147 & -    & -   & -107.820 & 15.669 & 10.21 \\
      \hline
    \end{tabular}
    \caption{Low energy parameters of the helium dimer, calculated with the
LM2M2 interaction, and those of the $np$ $1^+$ and
    $0^+$ states, and the $nn$ $0^+$ state calculated with the AV18
interaction. For the $0^+$ states the energy values are those of the virtual state.
Parameters of modified helium dimers, as explained in the text, are also shown. 
The values of the asymptotic normalization constant $C_a$ and of the mean square 
radius are also reported. See the main text for more details.}
    \label{tab:table1}
\end{table}

\subsection{Trajectories in the universal window}

The panel (b) of Fig.~\ref{fig:fig1} defines a Gaussian characterization of the unitary window. The position of real systems 
on the Gaussian curve identifies the characteristic ranges. The associated Gaussian potentials 
can be considered as a low-energy representation of the two-body interaction of the systems.
Through the variation of the Gaussian strength a system can (ideally) be moved  along the unitary window. 
Physical systems exist at their physical points, so the interaction has to be modified to move them from that point. At present, this can be done in the case of the residual interaction
between atoms by applying magnetic fields to change their electronic structure. 
The difficult technical implementations of this procedure could limit the knowledge of 
the new interaction allowing only to trace a few parameters of it. 
We refer for example to the sector of trapped cold atoms in which the applied magnetic field 
is related to changes in the two-body scattering length. When this parameter is allowed to take large values 
(and eventually diverges) the system moves inside the universal window. Since the window is characterized essentially by two parameters, $a_B$ and $a$, the lack of knowledge of the complete 
interaction is not important: the low-energy properties of the system inside the window are  
determined by them. Accordingly the characterization of the universal window by the 
Gaussian potential could be of interest. 

To analyze possible trajectories along the unitary window we use as example the
LM2M2 interaction of two helium atoms and define
\begin{equation}
V_\lambda = \lambda V_{LM2M2} \,.
\end{equation}
The value $\lambda=1$ refers to the original potential whereas for slightly bigger and lower
values of $\lambda$ the system moves along the window. The different $\lambda$
values generate fictitious helium dimers mimicking possible
modifications of the original potential. For selected cases of $\lambda$ the
corresponding low energy quantities are given in Table~\ref{tab:table1} and shown
in Fig.~\ref{fig:fig1}, panel (c), as solid triangles. The two-nucleon 
systems are shown on the curve too and, for the given values of $\lambda$, the
position of these modified helium dimers travel along the curve coinciding 
in specific cases with the nuclear systems. For each case
the range of the Gaussian potential that reproduces the values of $a_B$ and $a$  is
given in the last column of Table~\ref{tab:table1}. We notice that the Gaussian
range of the modified dimers varies very little along the window showing that it is possible to define
a characteristic Gaussian range associated to the helium dimer.

The above analysis is useful to characterize the universal behavior in terms of the
position of a system inside the universal window. Similar locations inside the universal
window imply similar dynamical properties. This can be put in evidence using the
wave function to calculate several observables, such as the mean square radius
\begin{equation}
    \langle r^2 \rangle = \frac{r_0^2}{4} \int_0^\infty dz\, z^2 \phi_B(z)^2
     = \frac{a^2}{8} (1+ \left(\frac{r_B}{a}\right)^2 +
o(\left(\frac{r_B}{a}\right)^3) ) \simeq \frac{a_B^2}{8} e^{2r_B/a_B} \,,
\end{equation}
and the asymptotic normalization constant, defined when
$\phi_B(z>2r_B/r_0)\rightarrow C_a e^{-zr_0/a_B}$ and directly related to the residue of the $S$-matrix at the momentum pole $k=i/a_B$
\begin{equation}
    C_a^2 \simeq \frac{2}{a_B}\;\frac{1}{1-r_e/a_B} =\frac{2}{a_B}
e^{2r_B/a_B}\,.
    \label{eq:ca2}
\end{equation}
The above quantities explicitly depend on the two low energy data and we have
introduced, valid up to third order, the scaling function
\begin{equation}
     f_{sc}= \frac{1}{1-r_e/a_B}=e^{2r_B/a_B}\,.
\label{eq:fsc}
\end{equation}

Inside the universal window observables are controlled by the large parameter $a_B$ 
with corrections given by the small parameter $r_e/a_B=2r_B/a$ or $\epsilon=2r_B/a_B$. 
In the case of $\langle r^2\rangle $ and $C^2_a$ this is encoded in the scaling function
$f_{sc}$ as explicitly shown in Fig.~\ref{fig:fig1}, panel (d).
The values given in Table~\ref{tab:table1}, properly divided by the indicated
factors, are located in the figure and result on top of the
scaling function at the corresponding value of $\epsilon$. 
The collapse on the curve is well verified for very different systems, in particular close to the unitary
limit. This analysis puts in evidence the CSI and the universal characteristic of the window.
Moreover, it shows that the dynamics is determined by the small parameter $\epsilon$ as continuously emerging
from the unitary point ($\epsilon=0$).

\subsection{Correlations inside the universal window}

When the interaction between two particles is strongly repulsive at short distances the two-body 
system is, as a consequence, highly correlated. For bound systems the probability to be inside the repulsive core 
is very small. Accordingly, the wave function in that region is almost zero and increases rapidly 
towards the attractive region. Therefore the total energy results from a big cancellation between 
the kinetic and potential energy. Systems such as the helium dimer or the deuteron are examples of this 
kind of correlation. It is interesting to analyze the description of these systems in terms of the 
low-energy parameters. Outside the interaction region 
the $s$-wave reduced wave function of the system is
\begin{equation}
    \psi_B(r\rightarrow\infty)= C_a e^{-r/a_B}
\end{equation}
and the probability $P_e$ to be in that region is defined as
\begin{equation}
    P_e=C_a^2 \int_{2r_B}^\infty e^{-2r/a_B} dr= C_a^2 \frac{a_B}{2} e^{-4 r_B/a_B}=
    \frac{1}{1-r_e/a_B} e^{-4r_B/a_B}=e^{-2r_B/a_B}=\frac{1}{f_{sc}} \, ,
\end{equation}
where we have used Eq.(\ref{eq:ca2}) and we have identified
$2r_B$ as the lower limit for two particles to be considered outside the
interaction region.
Accordingly $P_e$, the probability to be outside the interaction region,
is the inverse of the scaling function. 
For weakly bound systems this quantity is governed by the ratio
$2r_B/a_B$, therefore we consider the systems inside the unitary window as strongly
correlated.


\section{The three-body universal window}

In this section we discuss the three-body universal window for three equal bosons and three equal fermions 
with $1/2$ spin-isospin symmetry, of interest for nuclear physics, in terms of the Gaussian characterization.
The three-body system inside the window has remarkable properties as the Efimov effect,
a manifestation of the discrete scale invariance that strongly constrains
the three-body physics. The CSI of the two-body system is 
broken at the level of three particles by the introduction of a new scale governed by the energy value of 
the three-body system at unitarity. On one side, these interesting properties triggered an enormous amount 
of experimental work directed to study the behavior of three particles inside the universal window
\cite{kraemer:2006_Nature,grisenti:2000_Phys.Rev.Lett.,zaccanti:2009_NatPhys,berninger:2011_Phys.Rev.Lett.,machtey:2012_Phys.Rev.Lett.,roy:2013_Phys.Rev.Lett.,klauss:2017_Phys.Rev.Lett.}. 
On the other side it would be of fundamental importance to understand
correlations between low-energy properties and the specific location of a system inside the window. In particular, in
the case of nucleons, these correlations will be taken as signatures of universal behavior.

\subsection{The three-boson system: bound states}
In the case of a zero-range interaction the three-body system turns out to be unbound from below 
(Thomas collapse~\cite{thomas:1935_Phys.Rev.}). Its spectrum, deduced by V. Efimov, is given by the Efimov radial law~\cite{efimov:1970_Phys.Lett.B,efimov:1971_Sov.J.Nucl.Phys.}:
\begin{eqnarray}
    \frac{E_3^{(n)}}{E_2} =\tan^2\theta & \\
    E_3^{(n)}+ E_2 = & e^{-2(n-n_*)\pi/s_0}\, e^{\Delta(\theta)/s_0} E_* \,.
    \label{eq:zero3}
\end{eqnarray}
For each value of the angle $\theta$,
the binding energy of level $n$, $E_3^{(n)}$, is determined simultaneously by
the two-body binding energy, which in the zero-range limit ($a=a_B$) is $E_2=\hbar^2 / m a^2$, 
and by the binding energy of level $n_*$ at the unitary limit, $E_*=\hbar^2 \kappa_*^2/m$, 
defining the three-body parameter $\kappa_*$. The
function $\Delta(\theta)$
is a universal function, the same for all levels, governing the values of the three-body binding energy
inside the window. With the above definition, $\Delta(-\pi/2)=1$, and
parametrizations of the universal function exist
\cite{braaten:2006_PhysicsReports,naidon:2017_Rep.Prog.Phys.,gattobigio:2019_Few-BodySyst.}
for $\theta$ varying in the range $[-\pi,-\pi/4]$. At $\theta=-\pi/2$, $E_2=0$ 
and the spectrum shows the Efimov effect: a geometrical tower of states with constant energy ratios
$E_3^{(n)}/E_3^{(n+1)}=e^{2\pi/s_0}$ where, in the case of 
three-equal bosons, the universal number is $s_0 = 1.006237\ldots$.
The zero-range spectrum of Eq.(\ref{eq:zero3})
verifies a DSI. It results invariant when the scattering length $a$ is scaled by
the factor $e^{m\pi/s_0}$, with $m$ an integer number, maintaining invariant the three-body
parameter $\kappa_*$ and the angle $\theta$.

The zero-range model can be extended to consider the finite-range character of the interaction. In this case the
Thomas collapse is not present any more and the three-body spectrum 
can be written as
\begin{equation}
    E_3^{(n)}+ E_2= e^{\Delta_3^{(n)}(\theta)/s_0} E_*^{(n)}
    \label{eq:finite3}
\end{equation}
where $n=0,1,\ldots$ indicates the energy levels and $\Delta_3^{(n)}$ is the $n$-level function defined as
\begin{equation}
 \Delta_3^{(n)}(\theta)=s_0\log\frac{E_3^{(n)}+E_2}{E_*^{(n)}} \, .
\label{eq:deltan} 
\end{equation}
It depends on the particular interaction used to compute the energy values.
Moreover $E_2=\hbar^2/m a_B^2$ and $E_*^{(n)}=\hbar^2 [\kappa_*^{(n)}]^2/m$ is
the energy of level $n$ at the unitary limit, defining the three-body parameter 
of each level, $\kappa_*^{(n)}$. When finite-range potentials are used to
compute the $n$-level function the following behavior is verified~\cite{alvarez-rodriguez:2016_Phys.Rev.A} 
\begin{eqnarray}
& \Delta_3^{(n)}(\theta)  \rightarrow \Delta(\theta)  \hspace{0.5cm} & n>1  \\
& \frac{E_*^{(n)}}{E_*^{(n+1)}} \rightarrow e^{2\pi/s_0}  \hspace{0.5cm} & n>1
\,.
\end{eqnarray}
Only the lowest levels, and in particular the ground state ($n=0$), show range
effects. Starting from $n=2$
the energy spectrum closely tends to the zero-range spectrum of Eq.(\ref{eq:zero3}). 
The practical use of Eq.(\ref{eq:zero3}) and Eq.(\ref{eq:finite3}) depends on the knowledge of the 
universal or level functions $\Delta(\theta)$ or $\Delta_3^{(n)}(\theta)$ respectively. In the first case 
it is possible to solve the Skorniakov-Ter-Martirosian (STM) equations~\cite{kharchenko:1972_Sov.J.Nucl.Phys.} 
for different values of the 
two-body scattering length $a$ to cover the region of interest given by 
$-\pi<\theta<-\pi/4$~\cite{gattobigio:2019_Few-BodySyst.}. In the case of finite-range interactions the knowledge 
of the $n$-level function along the unitary window is related to the knowledge of the interaction in 
that region. In general the interaction is known at one point, the physical point, and to explore the unitary 
window some assumptions are needed. Many times scaled potentials have been used to slightly increase or reduce
their strength as a way to explore the universal window, here we use the Gaussian potential of 
Eq.(\ref{eq:tbg}) as the reference interaction to characterize the universal window.  

The results for a Gaussian potential of range $r_0$ with variable strength can be
summarized in the following
equations~\cite{kievsky:2013_Phys.Rev.A,kievsky:2014_Phys.Rev.A,gattobigio:2014_Phys.Rev.A,kievsky:2015_Phys.Rev.A}
\begin{eqnarray}
    & a_B \kappa_3^{(n)}&= \tan\theta \\
    & r_0\kappa_3^{(n)} &= \gamma_3^{(n)} {e^{\Delta_3^{(n)}(\theta)/2s_0}}{\sin\theta}\,,
    \label{eq:gauss3}
\end{eqnarray}
with $\gamma_3^{(n)}=r_0 \kappa_*^{(n)}$ and $E^{(n)}_3=\hbar^2 [\kappa_3^{(n)}]^2/m$.
The level functions $\Delta_3^{(n)}$ are computed
solving the Schr\"odinger equation with a Gaussian potential with variable
strength whereas the pure numbers
$r_0 \kappa_*^{(n)}=\gamma_3^{(n)}$, define the three-body parameter of each level at $\theta=-\pi/2$. 
It should be noticed that $\Delta_3^{(n)}$ and $\gamma_3^{(n)}$ are the same
for all Gaussian potentials. 

\begin{figure}[b]
    \centering
 \includegraphics[scale=0.40] {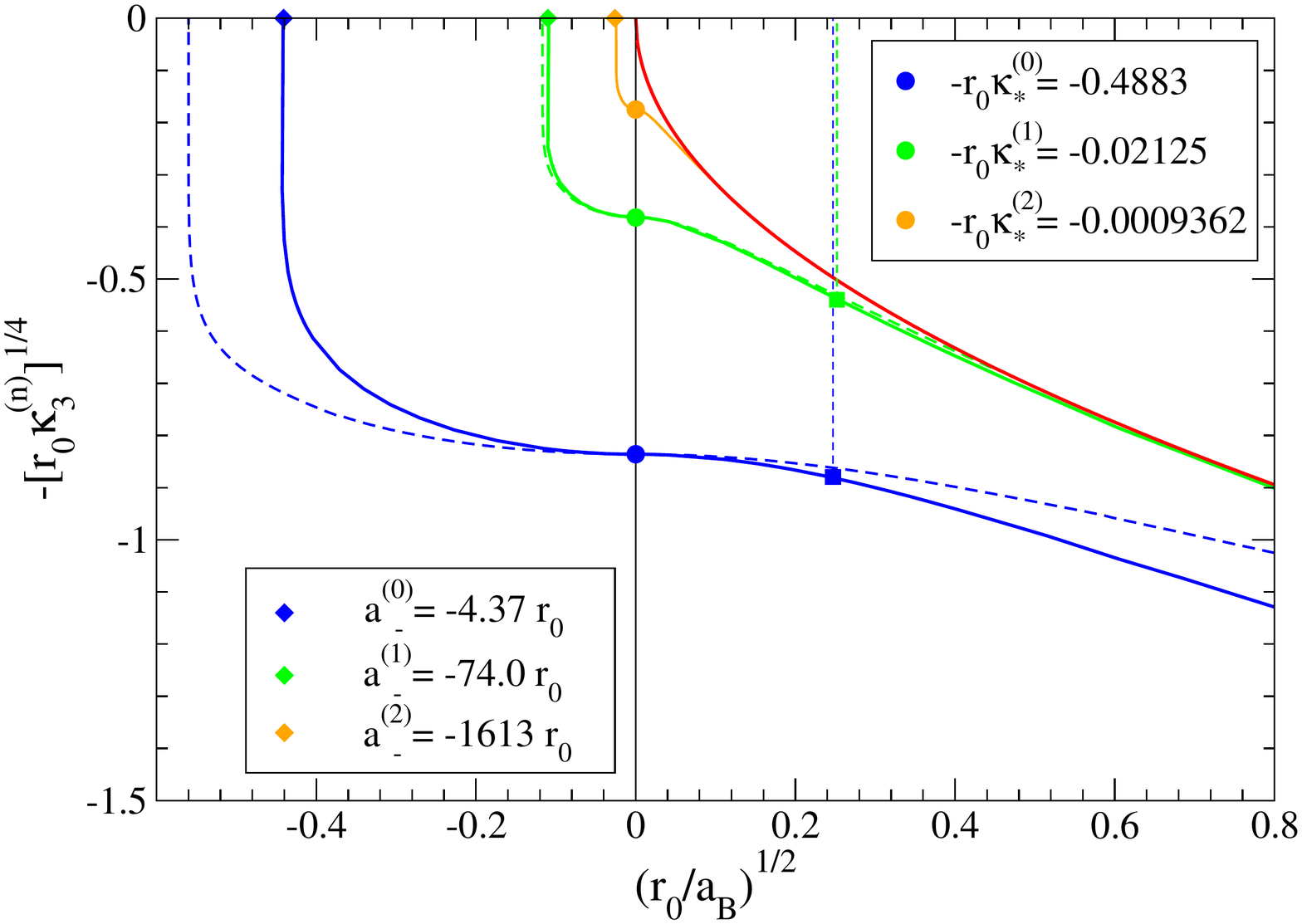} 
    \caption{The dimensionless quantity, $-[r_0\kappa_3^{(n)}]^{1/4}$, for
$n=0,1,3$, are shown as a function of $(r_0/a_B)^{1/2}$ for a Gaussian potential
(solid lines). The dashed lines are the results from the zero-range model. Notable values at
$\theta=-\pi/2$ and $-\pi$ are shown as solid circles and diamonds respectively.
The solid squares represent the two levels of the helium trimer on the Gaussian
$n=0,1$ levels.}
    \label{fig:fig3}
\end{figure}

In Fig.~\ref{fig:fig3} the first three levels of the Gaussian potential are shown
(solid lines) in a $[(r_0/a_B)^{1/2},-[r_0\kappa_3^{(n)}]^{1/4}]$ plot. The powers
$1/2$ and $1/4$ in the axis variables are used to make more visible the three levels. 
The zero-range results (dashed lines) are shown too, making a correspondence
of the two models at the unitary limit. Range effects are appreciable in the ground state, 
very reduced in the first excited state and are almost negligible in higher levels. The
solid circles on the $r_0/a_B=0$ axis indicates the first $\gamma_n$ values whereas the points where the
bound states disappear into the three-body continuum, $E^{(n)}_3=0$, are the
corresponding values of the scattering length
$a^{(n)}_-$ shown as solid diamonds, in units of the Gaussian range. 
Using those values the almost model independent quantities can be extracted
\begin{eqnarray}
\kappa_*^{(0)} a_-^{(0)}&=&-2.14  \\
\kappa_*^{(1)} a_-^{(1)}&=&-1.57  \\
\kappa_*^{(2)} a_-^{(2)}&=&-1.51 \, .
\end{eqnarray}
In the case of the ground state the estimate for van der Waals systems is 
$\kappa_*^{(0)} a_-^{(0)}\approx -2.2$ (see Ref.~\cite{naidon:2017_Rep.Prog.Phys.} 
and references therein). Therefore the Gaussian
characterization captures most of the ingredients of those systems inside the universal window.
Moreover, in the $n=1,2$ cases, the values tend rapidly to the zero-range
value of $a_-\kappa_*=1.507$~\cite{gogolin:2008_Phys.Rev.Lett.}.

In Fig.~\ref{fig:fig3} the two levels of the helium trimer using the LM2M2
interaction, $E_3^{(0)}=126.4\,$mK and $E_3^{(1)}=2.27\,$mK 
are shown as solid squares. Noticing that $E_2=1.303\,$mK, the position of these data on the plot are fixed
through the angle $\theta$ defined as $E_3^{(n)}/E_2=\tan^2\theta$.
The axis value of $r_0/a_B=0.061$, corresponding to the ground state, can be used to determine the 
characteristic Gaussian range $r^{(0)}_0=11.15\, a_0$ with which a Gaussian
potential reproduces the dimer and ground state trimer energies. From that value, the
three-body parameters of the helium trimer, ground and excited states, can be
estimated~\cite{deltuva:2020_Phys.Rev.C}
\begin{eqnarray}
    E_*^{(0)}= \frac{\hbar^2}{m} \left[\frac{\gamma_0}{r^{(0)}_0}\right]^2=83.1\, {\rm mK} \\
    E_*^{(1)}= \frac{\hbar^2}{m}
\left[\frac{\gamma_1}{r^{(0)}_0}\right]^2=0.157\, {\rm mK}
\end{eqnarray}
in complete agreement with the predictions given in literature~\cite{hiyama:2014_Phys.Rev.A,kievsky:2020_Phys.Rev.A}. 
Moreover at the three-atom continuum the characteristic
range predicts the value $a_-^{(0)}=-48.7\,a_0$ in
agreement with the helium values at that point, see Ref.~\cite{hiyama:2014_Phys.Rev.A}. Using
the scaled van der Waals length of helium, ${\tilde r}_{vdW}=\lambda^{1/4} r_{vdW}$, the Gaussian trajectory
predicts $a_-^{(0)}/{\tilde r}_{vdW}\approx-9.6$, in close agreement with
the universal value observed in van der Waals species, see
Ref.~\cite{naidon:2017_Rep.Prog.Phys.}
and references therein.

\subsection{The three-boson system: scattering states}

Considering three equal, spin 0, atoms as representative of the three-boson system, 
the Gaussian characterization of the universal window can be applied to study 
the atom-dimer scattering length $a_{AD}$. 
In the zero-range limit its expression, derived by Efimov~\cite{efimov:1979_Sov.J.Nucl.Phys.}, is
\begin{equation}
a_{AD}/a_B=d_1 +d_2\tan[s_0\ln(\kappa_*a_B)+d_3]\,,
\label{eq:a_ADB}
\end{equation}
where $d_1$, $d_2$ and $d_3$ are universal numbers and $\kappa_*$ is the
three-body parameter belonging to one of the three-body energy branches. The log-periodic functional form
of the observable is a consequence of the constraints imposed by the DSI. As $a_B\rightarrow\infty$, the
ratio $a_{AD}/a_B$ forms different branches with asymptotes located at values of
$a_B$ at which the three-body levels disappear into the atom-dimer continuum. 
In the case of finite-range interactions we use the parametrization proposed in
Ref.~\cite{kievsky:2013_Phys.Rev.A}
\begin{equation}
a_{AD}/a_B=d_1
+d_2\tan[s_0\ln(\kappa^{(n)}_*r_0(a_B/r_0)+\Gamma_3^{(n)})+d_3]\,,
\label{eq:a_ADR}
\end{equation}
where the pure number, $\kappa^{(n)}_* r_0=\gamma_3^{(n)}$, is used as the driving
term and we have introduced the finite-range three-body parameter $\Gamma_3^{(n)}$, as
discussed in Refs.~\cite{kievsky:2013_Phys.Rev.A,kievsky:2014_Phys.Rev.A}, to absorb finite-range corrections.

\begin{figure}[b]
    \centering
\includegraphics[scale=0.55] {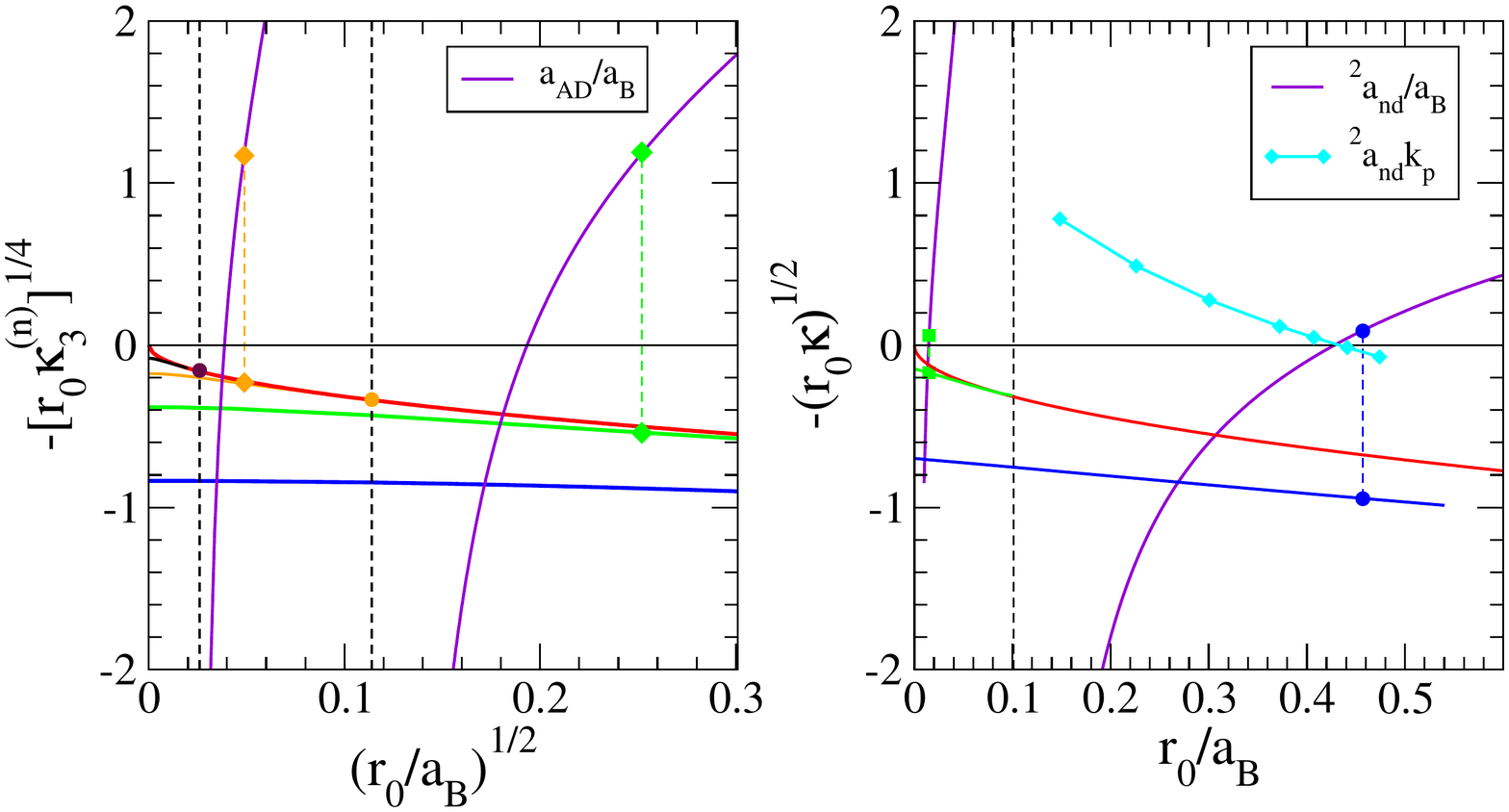}

    \caption{Left panel: The $a_{AD}/a_B$ function (violet solid line) inside
 the $(r_0/a_B)^{1/2},-[r_0\kappa_3^{(n)}]^{1/4}$ diagram, as explained in the text.
 Right panel: The $a_{nd}/a_B$ function (violet solid line) and the pole momentun $k_p$ (multiplied by $a_{nd}$) inside
 the $(r_0/a_B),-[r_0\kappa_3^{(n)}]^{1/2}$ diagram, as explained in the text.}
    \label{fig:fig4}
\end{figure}

We analyze the behavior of $a_{AD}$ inside the unitary window using a Gaussian
potential. Following Ref.~\cite{deltuva:2020_Phys.Rev.C},
we show in Fig.~\ref{fig:fig4} (left panel) two branches of the function 
$a_{AD}/a_B$ (violet solid line) using $\gamma_3^{(1)}=0.02125$ as driving term,
$d_1=1.541$, $d_2=-2.080$, $d_3=-2.038$ and $\Gamma_3^{(1)}=0.061$. 
In the figure the lowest four energy
levels are shown too: ground state (blue line), first  (green line),
second  (orange line) and third  (black line) excited states. The
two-body energy is represented by the solid red line and the two
asymptotes (dashed lines), extracted from Eq.(\ref{eq:a_ADR}) and located at $r_0/a_B=0.0139$ and $0.00059$,
indicate the positions at which the third (black solid circle) and second 
(orange solid circle) excited states disappear into the continuum. 
The positions on the
characteristic Gaussian curve of the first excited state of the helium trimer on the $n=1$ level
(lower green diamond) and on the $n=2$ level (lower orange diamond) are shown too.
These points correspond to the crossing of a straight line passing through the origin,
defined by the angle $E_3^{(1)}/E_2=\tan^2\theta$, with the $n=1,2$ levels.
In the case of the $n=1$ level, the value of the axis $r_0/a_B=0.0637$ corresponds
to the value $a_{AD}/a_B=1.19$ (higher green diamond).
Therefore the Gaussian characterization of the unitary window predicts
the atom-dimer scattering length to be $a_{AD}=1.19\, a_B$.
Using the LM2M2 value, $a_B=182.22\, a_0$, the value $a_{AD}=217 \, a_0$ is
obtained which has to be compared to the LM2M2 value for this quantity of 
$218.4\, a_0$~\cite{carbonell:2011_Comp.Rend.Phys.}. This demonstrates the capability of the Gaussian
characterization of the universal window to take into account accurately
finite-range effects. Accordingly, within an EFT we consider this result at the NLO level, in the sense that it includes range corrections.

The DSI allows to map the excited state of the trimer on a higher branch as
it is given by the lower orange diamond on the left panel of Fig.~\ref{fig:fig4},
corresponding to the values $r_0/a_B=0.00239$ and $a_{AD}/a_B=1.17$ (higher
orange diamond). The prediction is now $a_{AD}\approx 213 \, a_0$.
Within an EFT we consider this result as corresponding to the LO of the EFT, as the $n=2$ or higher branches have almost negligible
finite-range effects. This simple analysis shows the strong correlation existing
between low-energy observables inside the unitary window. Moreover it shows how the different branches can be 
used to estimate finite-range effects. Recent studies of the three-boson
continuum can be found in \cite{deltuva:2020_Phys.Rev.C1}

\subsection{The three nucleon system}

The two-nucleon system in states $J^\pi=0^+$ and $1^+$ belongs to the universal window. The $0^+$ state
is an $s$-wave state whereas the $1^+$ has a dominant $s$-wave component at low energies, in the case of
the deuteron it is about $95\%$.
The lightest nuclei, $^2$H, $^3$H, $^3$He and $^4$He have large probabilities to
be in $L=0$ and therefore we expect to observe universal properties.
Important questions to be clarified are the lack of excited states in the three-
and four-nucleon systems. Moreover the doublet neutron-deuteron scattering length, 
$^{2}a_{nd}\approx 0.65\,$fm has a very small value compared to the triplet 
neutron-proton scattering length $a_{np}\approx 5.2\,$ fm. In addition, data
for low energy neutron-deuteron scattering reveal the presence of a triton
virtual state. These properties can be traced back to the position of the 
nuclear system inside the universal window.

The study of the universal window in the case of three nucleons has to consider 
the two different values of the singlet and triplet scattering lengths, $a_s$ and $a_t$. 
Among different possibilities we choose to maintain the ratio $a_s/a_t$ close to
the experimental value, $a_s/a_t=-4.38$, in our
exploration of the unitary window~\cite{kievsky:2016_Few-BodySyst}. Therefore the 
change in one value fixes the value of the other. To characterize the universal window 
we construct a spin-dependent Gaussian potential 
with different strengths and ranges in the spin-isospin channels $S,T=0,1$ and $1,0$
\begin{equation}
    V(r)=V_0 e^{-r^2/r_0^2} {\cal P}_0 + V_1 e^{-r^2/r_1^2}  {\cal P}_1
    \label{eq:potf}
\end{equation}
where ${\cal P}_0$ projects onto the $S,T=0,1$ channel and
${\cal P}_1$ onto the $S,T=1,0$ channel. 
In the following we study the spectrum of the three-nucleon $J^\pi = 1/2^+$
state considering $r_0=r_1$, for which choice, 
at the unitary limit, the spectrum coincides with the boson case.
The Gaussian strengths are varied to examine the plane $(r_0/a_B,-r_0
\kappa^{(n)}_3)$, with $E^{(n)}_3=\hbar^2 [\kappa_3^{(n)}]^2/m$ being the binding energy of level $n$ 
and $E_2= \hbar^2/m a_B^2$ the two-body binding energy of the triplet state.
In Fig.~\ref{fig:fig4}, right panel, we show the ground state, $n=0$ (blue line) and first excited state, 
$n=1$ (green line) of the $J=1/2^+$ three-nucleon system whereas the red line is the ground state of 
the $1^+$ two-nucleon system.  
The $^3$H nucleus is mapped on the Gaussian ground state curve as a  blue circle
at coordinates verifying $\kappa^{(0)}_3 a_B=\tan\theta=1.95$ corresponding to the
square root of the ratio of the triton binding energy of $8.48\,$MeV with the
deuteron binding energy of $2.224\,$MeV. At that point $r_0/a_B=0.457$ from which
the characteristic Gaussian range $r^{(0)}_0=1.97\,$fm can be estimated and used 
to assign a value of the three-nucleon system at unitarity through the quantity
$\kappa_*^{(0)} r_0=0.4883$. We obtain $E_*^{(0)}\approx 2.55\,$MeV in good agreement with
previous estimates~\cite{epelbaum:2006_Eur.Phys.J.C,gattobigio:2019_Phys.Rev.C}.

Furthermore, we also show in the right panel of Fig.~\ref{fig:fig4} the doublet neutron-deuteron scattering length,$^2a_{nd}$, 
calculated with the Gaussian interaction, as the violet curve (in units of the energy
length $a_B$). It corresponds to a fit of the numerical results using the form given by Eq.(\ref{eq:a_ADR}), see Ref.~\cite{deltuva:2020_Phys.Rev.C}.
Two branches are shown, with the dashed vertical line, the asymptote at $r_0/a_B=0.101$, indicating
the position at which ${}^2a_{nd}$ diverges and the first excited state disappears into the
$1+2$ continuum. Using the characteristic range $r^{(0)}_0=1.97\,$ fm, we 
estimate the evaporation of the first excited state at $a_B=18.8\,$fm, corresponding
to a deuteron energy of around $0.12\,$MeV and a scattering length around $20\,$ fm,
very far from the corresponding physical values. This simple analysis explains
the one level structure of $^3$H in terms of its position inside the Gaussian
characterization of the unitary window.
The correlation between the ground state and the doublet scattering length can
be studied looking at the value of $^2a_{nd}/a_B$ for $r_0/a_B=0.457$, the
coordinate of the physical point on the ground state curve. This gives
${}^2a_{nd}/a_B=0.08$, indicated as the upper blue solid circle in the
figure. Using the deuteron energy length $a_B=4.32\,$fm, the resulting doublet scattering
length is $^2a_{nd}\approx0.4\,$fm. This value is slightly lower than the
experimental value of $0.65\,$fm, however this analysis explains the very low
value of this quantity if compared to the value of the $np$ triplet scattering
length. We observe the very delicate region in which $^2a_{nd}$ is located, 
where slightly different values of $r_0/a_B$ could produce large variations of
${}^2a_{nd}$, including a change of sign. The Gaussian characterization maps
${}^2a_{nd}$ in the correct (positive) region clarifying the strong correlation
between this quantity and the $^3$H energy, a property observed already many years 
ago~\cite{phillips:1977_Rep.Prog.Phys.}.
It is possible to use the higher branch of the ${}^2a_{nd}/a_B$ curve to determine the size of
finite-range corrections. The triton point is located on the $n=1$
level (lower green circle) at $r_0/a_B=0.015$ corresponding to ${}^2a_{nd}/a_B=0.06$, 
slightly lower than the value obtained analyzing the $n=0$ level.  As for the
boson case, these two estimates can be considered in the EFT as corresponding to the NLO and LO respectively.
This simple analysis explains some peculiarities of the nuclear system
strictly correlated to its location inside the universal window.

Finally we discuss the evolution of the three-nucleon virtual state after the $n=1$ level crosses the $1+2$ continuum. Following Refs.~\cite{deltuva:2020_Phys.Rev.C,rakityansky:2007_J.Phys.A:Math.Theor.} the $S$-matrix energy pole, $E_P=-3\hbar^2 k^2_p/4m$ is determined from the $s$-wave low energy phase-shifts calculated using the Gaussian potential of Eq.(\ref{eq:potf}). The behavior of the $a_{nd}k_p$ function is shown in Fig.~\ref{fig:fig4} (right panel) as a cyan solid line fitting the numerical calculations (cyan diamonds). This function crosses the physical point at $r_0/a_B=0.457$ from which the triton virtual state, $E_p=0.48\,$MeV can be extracted. The Gaussian characterization explains this value in agreement to experimental
determinations and theoretical investigations~\cite{girard:1979_Phys.Rev.C,orlov:2006_Phys.At.Nucl.,babenko:2008_Phys.At.Nucl.,adhikari:1983_Phys.Rev.C,yamashita:2008_Phys.Lett.B,rupak:2019_Phys.Lett.B}.

\section{Characterization of the unitary window for more than three particles}
The Gaussian characterization of the universal window can be extended to
describe systems composed by more than three particles. The DSI, which emerges
in the three-body sector and gives rise to the Efimov spectrum, strongly  constrains the $N>3$ (bosons) 
or $A>3$ (nucleons) energy spectrum.  For equal bosons, where the spatial wave function is symmetric, 
DSI can be observed well beyond three particles. In the case of $A$ nucleons, the spatial-symmetric wave function 
is dominant only up to four particles, and deviations from the 
bosonic-Efimov scenario appear for the $A>4$ levels; in this case 
it is interesting to explore how the energy levels emerge 
receding from the unitary limit. 

\subsection{The $N$-boson systems}
\begin{figure}
 \includegraphics[width=0.7\linewidth] {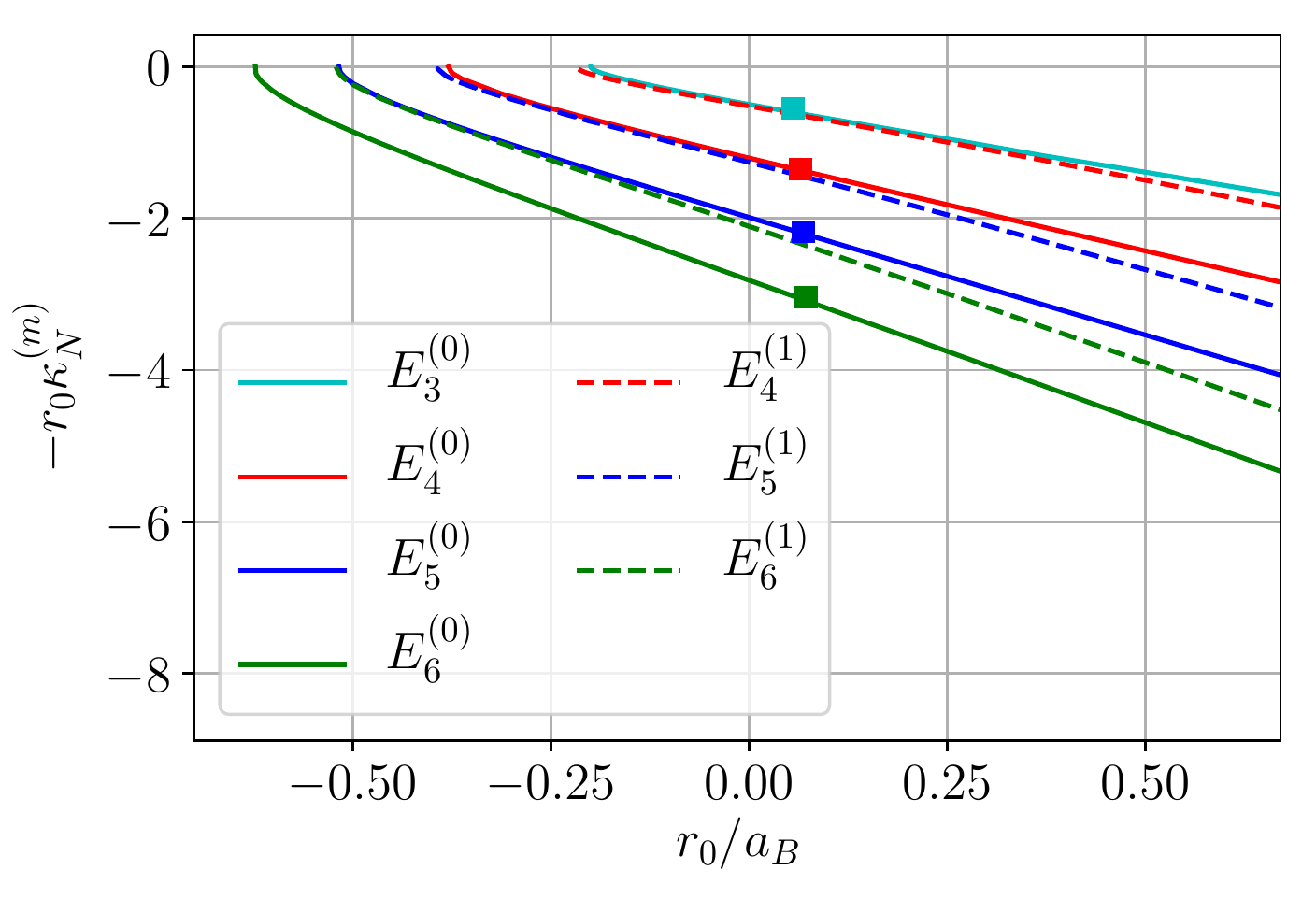}
 \caption{$N\le6$ energy levels given in terms of the energy momenta of the ground
 and excited states as a function of the inverse
 energy length $a_B$, both in units or $r_0$. The squares represent the 
 ground states of the $^4$He$_N$ clusters calculated with the 
 realistic TTY potential ~\cite{lewerenz:1997_J.Chem.Phys.}.}
 \label{fig:fewEfimov}
\end{figure}

The unitary window for $N$ bosons can be characterized using the Gaussian potential of Eq.~(\ref{eq:tbg}).
Tuning the strength of the potential the ground- and excited-state energies, 
$E_N^{(0)}$ and $E_N^{(1)}$, are calculated as a function of the two-body scattering
length $a$, or equivalently the energy length $a_B$. The results of these calculations are presented in
Fig.~\ref{fig:fewEfimov} for $N=4,5,6$. Results for $E_N^{(0)}$,
up to $N=70$, can be found in Ref.~\cite{kievsky:2020_Phys.Rev.A}.
A striking feature of the $N=4,5,6$ spectrum is the appearance of a
twin-level structure; for a given $N$, there are two bound states, one deep and one shallow, 
below each $N-1$ ground state. This pattern is expected to repeat itself for each 
three-body Efimov state, if the DSI is maintained, appearing as resonances 
in the $N$-body system. Studies in the four-body system exists~\cite{deltuva:2010_Phys.Rev.Aa,gattobigio:2011_Phys.Rev.A,gattobigio:2012_Phys.Rev.A,deltuva:2012_Phys.Rev.A,deltuva:2013_Few-BodySyst,greene:2010_Phys.Today}. 
The existence of the twin-level structure is not restricted to a number of particles $N\le6$; for the Gaussian
potential, the pattern is maintained up to
$N=12$~\cite{kievsky:2014_Phys.Rev.A,kievsky:2014_Few-BodySystems}. For a  
number of particles $N>12$ a third level appears as one consequence of the finite range character of the force.
The DSI smears out allowing for a transition between universal and non-universal behavior as the study of the unitary window is extended to consider deep bound states~\cite{kievsky:2020_Phys.Rev.A}. Limiting the discussion to the two-level structure, Eq.(\ref{eq:gauss3}) is extended for $N>3$ as
\begin{eqnarray}
    a_B \kappa_N^{(m)} & = & \tan\theta \\
    r_0\kappa_N^{(m)} & = & \gamma_N^{(m)} {e^{\Delta_N^{(m)}(\theta)/2s_0}}{\sin\theta}
\hspace{1cm}  m=0,1
    \label{eq:gaussn}
\end{eqnarray}
with $m=0$ being the $N$-body ground state and $m=1$ the excited state
close to the $(N-1)$-body threshold. The pure numbers
$\gamma_N^{(m)}=r_0\kappa_{*,N}^{(m)}$, determining 
the energies at the unitary limit, $E^{(m)}_{*,N}$, are
characteristic of every  Gaussian potential and their values, 
up to $N=6$, are given in Table~\ref{tab:table2}. 
The energy of the level $m$ is $E_N^{(m)}=\hbar^2 [\kappa_N^{(m)}]^2/m$
and $\Delta^N_m(\theta)$ is the Gaussian level function for $N$
bosons in the states $m=0,1$:
\begin{equation}
 \Delta_N^{(m)}(\theta)=s_0\log\frac{E_N^{(m)}+E_2}{E_{*,N}^{(m)}} \, .
\label{eq:deltam} 
\end{equation}
In the $N=4$ case, $\Delta_4^{(m)}(\theta)$ is explicitly given in Ref.~\cite{alvarez-rodriguez:2016_Phys.Rev.A} 
where it is compared to the zero-range four-body universal
function. To put in evidence the DSI character of the $N$-boson system, 
many efforts have 
been done to determine universal ratios between 
the $N$-body bound state energies at the 
unitary point in the limit of zero-range 
interaction. Precise numbers exist for $N=4$~\cite{deltuva:2010_Phys.Rev.Aa} whereas
estimates exist for higher systems
~\cite{hammer:2007_Eur.Phys.J.A,bazak:2016_Phys.Rev.A,dincao:2009_Phys.Rev.Lett.,vonstecher:2009_NatPhys,vonstecher:2010_J.Phys.B:At.Mol.Opt.Phys.}.
The Gaussian ratios $\gamma^{(0)}_N/\gamma^{(1)}_N$ and $\gamma^{(m)}_{N+1}/\gamma^{(m)}_N$
can be inferred from the values in Table~\ref{tab:table2}.

As illustration of the effectiveness of the Gaussian characterization we
analyze $^4$He$_N$ clusters, largely studied with realistic
helium-helium interactions~\cite{pandharipande:1983_Phys.Rev.Lett.,lewerenz:1997_J.Chem.Phys.}. 
We map these systems on the Gaussian curves of
Fig.~\ref{fig:fewEfimov} (solid squares) using the energies
calculated in Ref.~\cite{lewerenz:1997_J.Chem.Phys.}.
The position on the Gaussian curve fixes the ground state characteristic radius $r_N^{(0)}$ 
for each $N$-body system; it can be used to predict the energy of the ground and
excited state of the clusters at the unitary limit. The corresponding results
are given in Table~\ref{tab:table2}. For the sake of comparison, the results of He-He potential HFD-HE2 \cite{aziz:1979_J.Chem.Phys.}, re-scaled at the unitary limit as discussed in Ref.~\cite{kievsky:2020_Phys.Rev.A}, are shown in the last column. A remarkable agreement, better than $2\%$, is obtained.

In general, the knowledge of the $N=2-6$ energy values of a system belonging to
the universal window allows to construct a low-energy representation of the interaction 
that can be used to predict the ground state energy per particle, $E^{(0)}_N/N$, of the homogeneous system.
A strict correlation between the low-energy dynamics of the few-body
system and the many-body system, induced from the position of the system inside
the universal window,
exists~\cite{kievsky:2020_Phys.Rev.A,kievsky:2017_Phys.Rev.A,vankolck:2017_Few-BodySyst.,tan:2008_Ann.Phys.,tan:2008_Ann.Phys.a,weiss:2015_Phys.Rev.Lett.,weiss:2017_Few-BodySyst}. 

\begin{table}[]
    \centering
    \begin{tabular}{c|ccccc|c}
\hline
N & $\gamma^{(0)}_N$ & $\gamma^{(1)}_N$ & $r_N^{(0)}$   & $E^{(0)}_{*,N}$ & $E^{(1)}_{*,N}$ & $E^{(0)}_{*,N}$(HFD-HE2) \\
 4 & 1.1847 & 0.512 & 11.85~$a_0$ & 0.433~K & 0.081~K & 0.440 K \\
 5 & 1.955  & 1.240 & 12.50~$a_0$ & 1.059~K & 0.426~K & 1.076 K \\
 6 & 2.770 & 2.067  & 13.13~$a_0$ & 1.926~K & 1.073~K & 1.946 K \\
\hline
 4 & 1.1847 & 0.512 & 2.078~fm & 13.47~MeV  & 2.52~MeV &   \\
    \end{tabular}
 \caption{Gaussian pure numbers $r_0\kappa_{*,N}^{(m)}=\gamma_N^{(m)}$, $m=0,1$ at
unitarity, the characteristic range and the energies at the unitary limit for
bosonic helium (first rows) and four nucleons (last row). In the boson case, the results using the re-scaled HFD-HE2 potential at the unitary limit are shown in the last column.}
    \label{tab:table2}
\end{table}

\subsection{Collapse of finite-range interactions onto the zero-range model}
\begin{figure}
  \begin{minipage}{0.5\textwidth}
    \includegraphics[width=\linewidth]{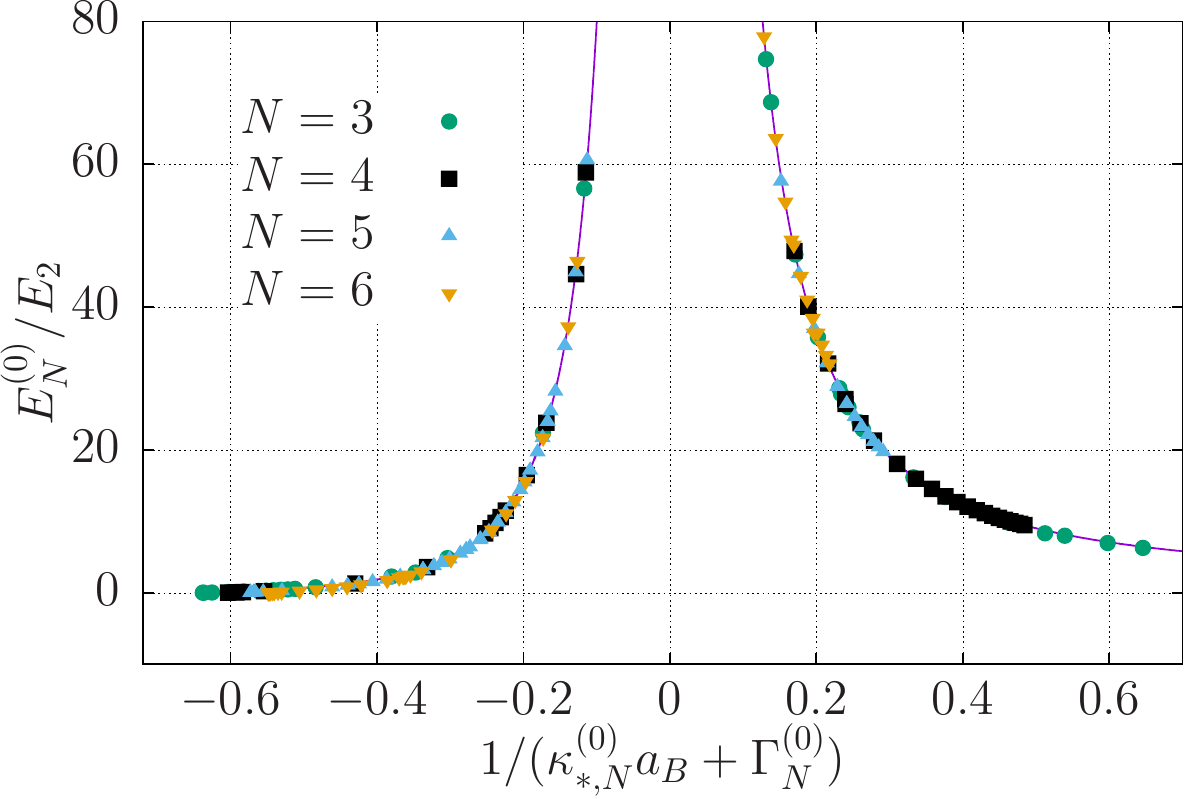}
  \end{minipage}
  \begin{minipage}{0.5\textwidth}
    \includegraphics[width=\linewidth]{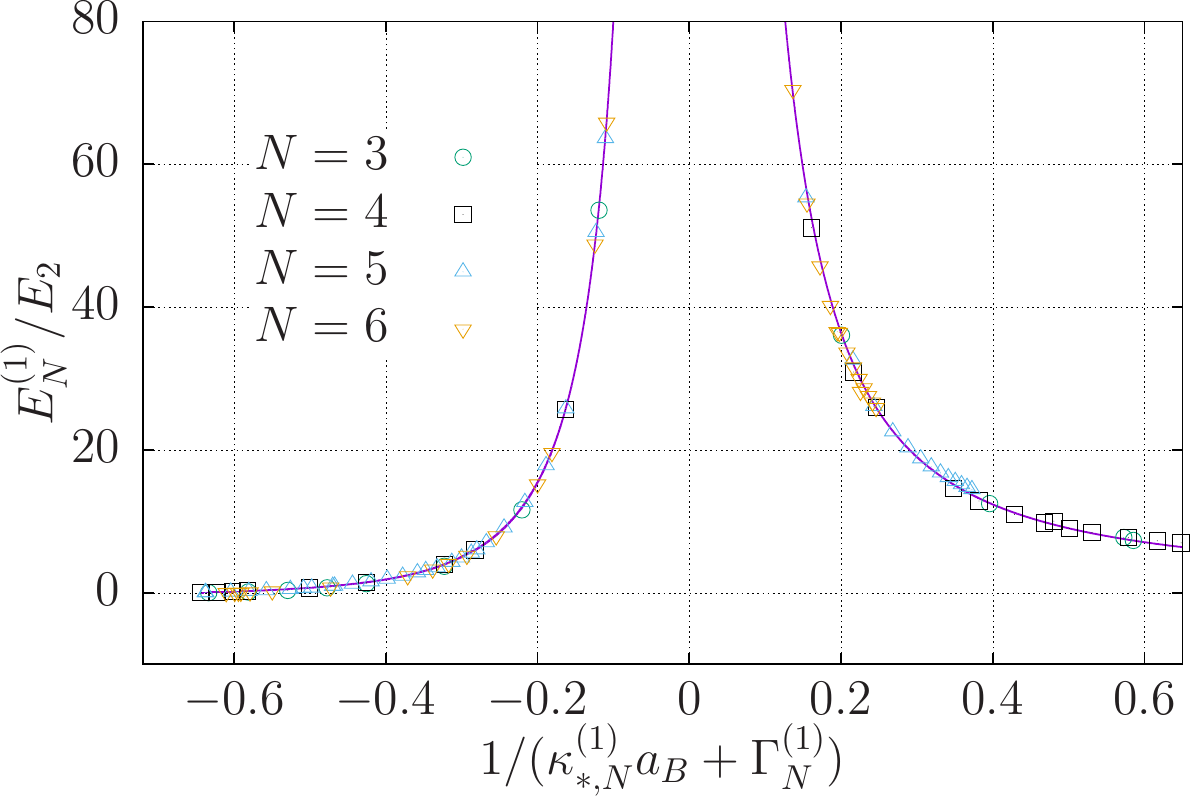}
  \end{minipage}
 \caption{The $N$ boson ground (left panel) and first excited (right panel) binding energies,
$E_N^{(0)}$ and $E_N^{(1)}$, in units of $E_2$ as a function of the inverse of the
energy length $a_B$, in units of the $N$-body parameter $\kappa_{*,N}^{(m)}$ shifted
by the finite-range parameter $\Gamma_N^{(m)}$. Different symbols
represent different potential models as described in
Ref.~\cite{gattobigio:2014_Phys.Rev.A}. Adapted from Ref.\cite{gattobigio:2014_Phys.Rev.A} with permission. }
 \label{fig:universality}
\end{figure}

We make one more step in the study of universal behavior of real systems located
inside the unitary window showing that the curves describing the $N$-body energies
as a function of the energy length for different number of particles are
actually the same curve. This is a manifestation of the strong constraints
imposed by the DSI and controlled by the three-body
parameter~\cite{vankolck:2017_Few-BodySyst.}.
Following Refs.~\cite{kievsky:2013_Phys.Rev.A,garrido:2013_Phys.Rev.A,gattobigio:2014_Phys.Rev.A} the Efimov
radial law, extended in Eqs.(\ref{eq:gauss3}) and (\ref{eq:gaussn}) to describe finite-range
interactions, can be related to the three-body universal function by the
introduction of the $N$-body finite-range parameter $\Gamma_N^{(n)}$ at different branches.
In the specific case of $N=3$, Eq.(\ref{eq:gauss3}) is modified by explicitly relating the finite-range spectrum to the zero-range universal function
$\Delta(\theta)$, as follows
\begin{equation}
  \kappa_3^{(n)} a_B = \tan\theta\,, \quad \kappa^{(n)}_* a_B + \Gamma_3^{(n)} =
  \frac{e^{-\Delta(\theta)/2s_0}}{\cos\theta}\,,
  \label{eq:finiteRangeEfimov}
\end{equation}
  whose origin has been traced to the running of the three-body
scale~\cite{gattobigio:2019_Few-BodySyst.,ji:2015_Phys.Rev.A}. What
 is unraveled is that each $N$-body system, ground and excited state,
has its finite-range parameter $\Gamma_N^{(0)}$ and $\Gamma_N^{(1)}$, 
so that  Eq.(\ref{eq:gaussn}) is modified as
\begin{equation}
  \kappa_N^{(m)} a_B = \tan\theta\,, \quad \kappa_{*,N}^{(m)} a_B + \Gamma_N^{(m)} =
  \frac{e^{-\Delta(\theta)/2s_0}}{\cos\theta}\,,
  \label{eq:fewEfimov}
\end{equation}
explicitly relating the description of the ground $m=0$ and the excited $m=1$ states
to the three-body universal function.
In Fig.~\ref{fig:universality} we see that the finite-range parameter
$\Gamma_N^{(m)}$, which encodes the finite-range corrections, can be used 
to make both, the ground and excited states of the few-body systems, collapsing on the 
three-body universal curve given by the Efimov radial law in Eq.(\ref{eq:zero3}). 
This is a clear sign that these systems belong
to the same universality class and that their spectra are constrained by a DSI 
governed by the three-body parameter $\kappa_*^{(0)}$. 

\subsection{The $A\le 6$ universal window}
\begin{figure}
  \begin{minipage}{0.5\textwidth}
    \includegraphics[width=\linewidth]{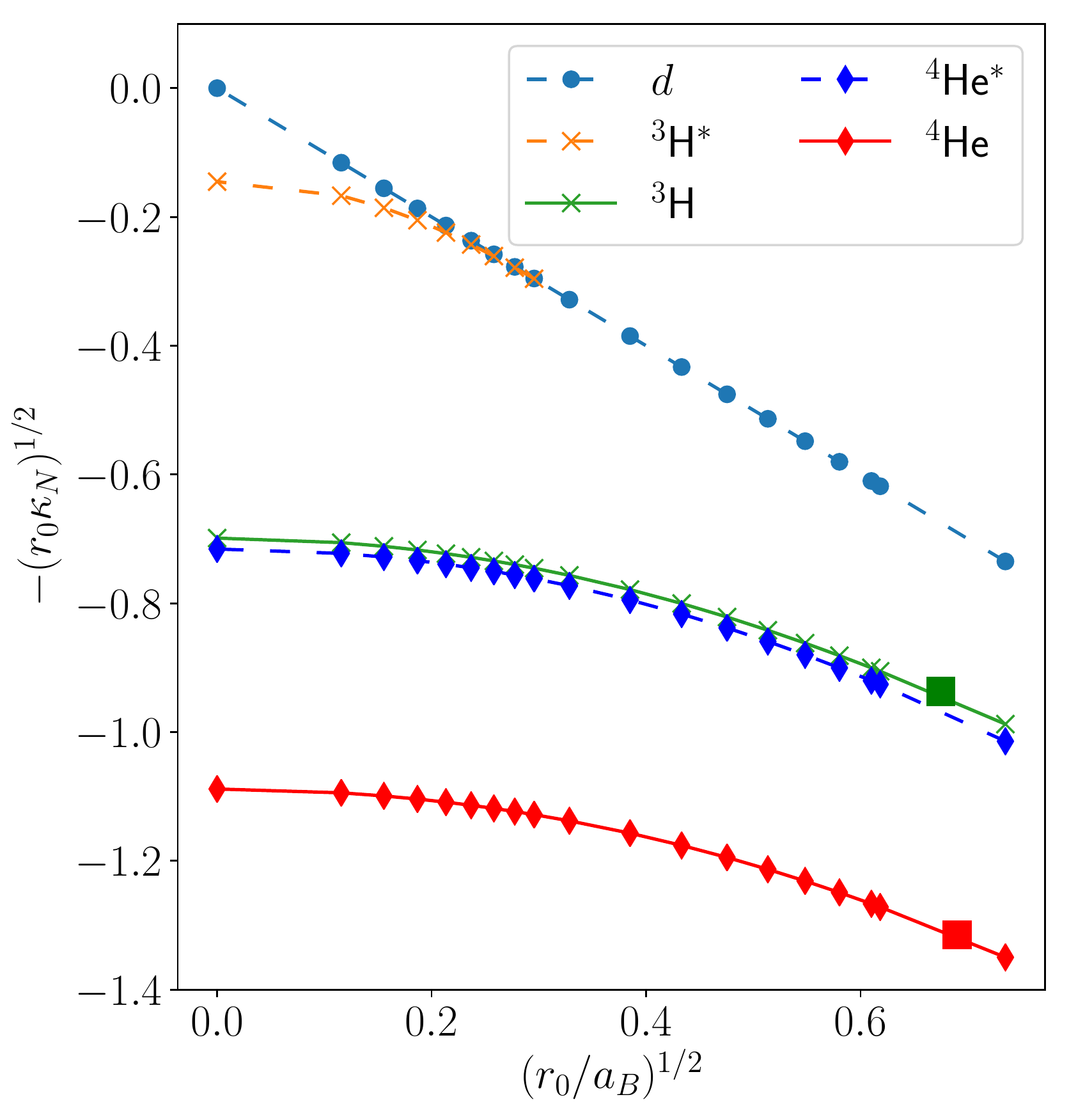}
  \end{minipage}
  \begin{minipage}{0.5\textwidth}
    \includegraphics[width=\linewidth]{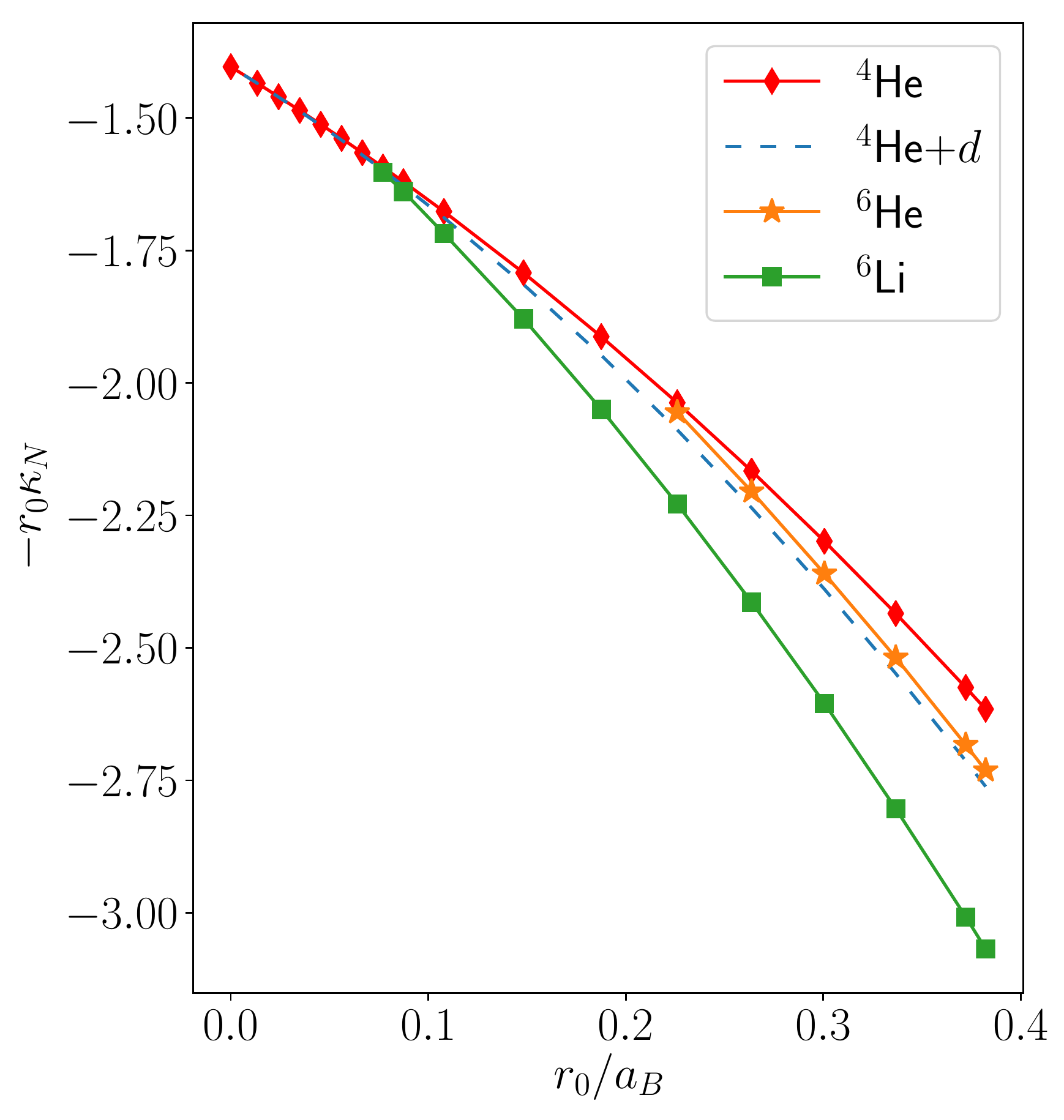}
  \end{minipage}
  \caption{Left panel: Square root of the binding momenta $\kappa_2$, $\kappa_3^{(n)}$, $\kappa_4^{(m)}$ 
for $d$, $^3$H, and $^4$He respectively along the nuclear
  cut $a_s/a_t=-4.3$ in terms of the inverse of the energy length $a_B$, both
  in units of the Gaussian range $r_0$. The position of $^3$H and $^4$He at the
physical point are given as a green and red solid squares respectively.
   Right panel: Binding momentum along the nuclear cut for $A=6$
   as a function of $a_B$ both in units of $r_0$. 
   In both panels the Coulomb interaction has not been taken into account. Adapted from Ref.\cite{gattobigio:2019_Phys.Rev.C} with permission.}
  \label{fig:efimovPlot}
\end{figure}

We have already discussed the universal character of the three-nucleon low energy
spectrum showing the existence of strong correlations between observables related to the
position of the two- and three-nucleon systems inside the universal window.
These properties suggest the possibility of describing nuclear physics as
continuously linked to the unitary limit~\cite{konig:2017_Phys.Rev.Lett.,platter:2005_PhysicsLettersB,gattobigio:2011_Phys.Rev.C,
hammer:2018_Few-BodySyst.}. 
Here we show the $A=4,6$, $L=0$, nuclear spectrum along the nuclear cut
$a_s/a_t=-4.38$ using the Gaussian two-channel potential of Eq.~(\ref{eq:potf}).
First of all, not considering the Coulomb interaction,
the $A=2-4$ spectrum is reported in Fig.~\ref{fig:efimovPlot}, left panel. The two three-nucleon states, 
$^3$H and $^3$He, are degenerate, moreover there is an infinite tower of excited states at unitarity as in the boson case. 
As the value of $r_0/a_B$ increases the excited states disappear one by one and the last one, indicated in the figure as 
$^3$H$^*$, disappears at $r_0/a_B=0.101$ resulting in the observed one level structure of $^3$H.
The four body spectrum has similar behaviour to the bosonic case: it has a two-level
structure, a deep state corresponding to $^4$He, and one excited state, $^4$He$^*$, close to the three-body threshold.
To be noticed that this state, which is a resonance, results bound without considering the Coulomb
interaction \cite{gattobigio:2019_Phys.Rev.C}.
The $^3$H and $^4$He nuclei can be mapped on the Gaussian curves through the
angles defined by the corresponding energy ratios, $E_3/E_2$ and $E_4/E_2$. 
They are indicated in Fig.~\ref{fig:efimovPlot}, left panel, as a green solid square ($^3$H) and as a red solid
square ($^4$He).
In the $^4$He case it should be taken $E_4=29.1\,$MeV, without considering the Coulomb
contribution~\cite{pudliner:1995_Phys.Rev.Lett.}. Its position on the plot
corresponds to a characteristic Gaussian range, $r^{(0)}=2.078\,$fm, from which
the binding energies at unitarity, $E^{(0)}_{*,4}$ and
$E^{(1)}_{*,4}$, can be deduced. They are reported on Table~\ref{tab:table2}.

The spectrum of the $A=6$ systems along the nuclear cut is reported in the right panel of
Fig.~\ref{fig:efimovPlot}. There are two different $A=6$ states 
discriminated by their spin-isospin quantum numbers: the $^6$He with $S=0,T=1$, and
the $^6$Li with $S=1,T=0$. Interestingly, neither are present at the
unitary limit, being above the corresponding thresholds
$^4$He and $^4$He + $d$, respectively. As $r_0/a_B$ moves toward positive values
they emerge from their thresholds,
first $^6$Li at $r_0/a_B=0.07$, and then $^6$He at $r_0/a_B=0.19$.
At the physical point, 
the light
nuclear spectrum (without considering the Coulomb interaction) consists in
one level for $^3$H and $^3$He, which are degenerated, two levels for $^4$He
and one level for $^6$He and for $^6$Li. The evolution of the excited $^4$He state
considering the Coulomb interaction is discussed in section 5.2. The present
analysis gives a simple explanation of the 
light nuclear spectrum as emerging continuously from the unitary limit.

\section{Implications of Efimov physics in determining the nuclear EFT}

The Gaussian characterization of physical systems in the Efimov window corresponds to a regularized version of the LO EFT description, where the Gaussian range is the inverse of the ultraviolet cutoff. In this formulation, finite range effects are implicitly contained in the cutoff and disappear as the latter is removed, recovering a scale invariant description. Even in this limit, a scale has nevertheless to be introduced at the 3-body level in the form of a dimensionful 3-body parameter \cite{braaten:2006_PhysicsReports,naidon:2017_Rep.Prog.Phys.}: as a matter of fact, the short-distance two-body dynamics does not decouple in the 3-body sector and manifests itself as an additional 3-body interaction in the LO EFT, designed to absorb all the cutoff dependence in the zero range limit. By specifying the corresponding strength through a 3-body datum, the continuous scale invariance is broken to a discrete scale invariance.

The sensitivity of the 3-body system to the short distance two-body dynamics depends solely on the proximity to the unitary limit, and persists after the inclusion of finite range effects \cite{kievsky:2017_Phys.Rev.C}. In particular, it also applies if the EFT is interpreted as a finite cutoff effective theory \`a la Lepage \cite{lepage:1997_Nucl.Phys.Proc.VIIIJorgeAndreSwiecaSummerSch.1995Ed.CBertulaniAl,Lepage:1989_TASI}, where the renormalization is done implicitly, through the fitting of low-energy constants, and cutoff independence is only attained up to neglected higher orders.
With a finite cutoff the Thomas collapse is avoided and there is a well defined 3-body ground state as well as, close to unitarity, all the higher Efimov states, as exemplified in Fig.~\ref{fig:fig3}.
In this perspective the cutoff is interpreted as a physical parameter related to the intrinsic scale of the theory, and therefore it is bound to assume values inside a given natural range. This constraint also identifies the spectrum of 3-body bound states as a function of the 2-body scattering length, or alternatively of the 2-body binding energy. Variations of the cutoff within the natural range induce drastic changes in the 3-body spectrum, the more so the closest the system is to the unitary limit, due to the existence of densely spaced Efimov states. Thus, close to the unitary limit, the extreme sensitivity to the cutoff also affects the finite-cutoff theory, because of the strong correlations between the ground state and the other bound states, reflecting the remaining DSI.
The sensitivity also concerns the continuum states, as exemplified in Fig.~\ref{fig:fig4}, where one can verify that small changes in $r_0$ produce a change of sign in the scattering length. 
The introduction of a LO 3-body force allows to set correctly the 3-body ground state energy, bringing the lowest branch of the Efimov plot to the curve that follows the evolution of the physical state to the unitary limit.

Stated differently, one can say that, without a 3-body force at LO, the scale of the 3-body ground state is a cutoff effect, and as such it is affected by a sizeable uncertainty. Close to the unitary limit, this uncertainty would propagate to all the tower of Efimov states, resulting in a very poor description of the shallowest ones. Thus, the EFT would be totally unable to describe those states which should, on the contrary, better fit in the domain of applicability of the EFT.
The intrinsic length scale of the underlying interaction can be reconstructed by locating the systems on the universal curves through the value of the corresponding Gaussian range $r_0$.
For example, a $N$-body physical system, having energy $E^{(0)}_N$, is mapped on the Gaussian characterization 
of the universal window through the energy ratio $E_N^{(0)}/E_2$, where $E_2$ is the energy of the
corresponding two-body system. Limiting the discussion to equal particles and
a single two-body energy level, this procedure is unambiguous. The position of
the system fixes the characteristic radius $r_N^{(0)}$ with which a
Gaussian potential with variable strength describes a path linking the physical
point, determined by $(E_2,E_N^{(0)})$, to the unitary point, determined by 
$E_2=0$ and $E_{*,N}^{(0)}=[\gamma_N^{(0)}]^2\hbar^2/m [r_N^{(0)}]^2$. Considering
different values of $N$ of the same physical system, different characteristic
ranges are obtained as it is clear on
Figs.~\ref{fig:fewEfimov},\ref{fig:efimovPlot}.
Though these different Gaussian potentials are useful to determine the paths to
the unitary limit, they define different potentials in each $N$-body sector, with different ranges all having the same order of magnitude. 

By introducing a 3-body force at the LO all these different descriptions can be unified as deriving from a single underlying effective Lagrangian, comprising two- and three-body contact interactions \cite{kievsky:2017_Phys.Rev.C}.

\subsection{The nuclear physical point}

The great complexity of QCD interactions produces very disparate phenomena at various scales.
In the chiral limit, spontaneous chiral symmetry breaking takes place, leading to the emergence of long-range collective modes, the Goldstone bosons, represented by the pions. Thus the chiral limit defines a critical point. Since chiral symmetry is only approximate, the pions acquire a mass but they keep their Goldstone bosons' character in that their interactions are weak at low energies. This  enables in turn the perturbative approach to nuclear interactions known as the Chiral-EFT or ChEFT \cite{vankolck:1999_Prog.Part.Nucl.Phys.,bedaque:2002_Ann.Rev.Nucl.Part.Sci.,epelbaum:2006_Prog.Part.NuclearPhys.,epelbaum:2009_Rev.Mod.Phys.,machleidt:2011_PhysicsReports,hammer:2020_Rev.Mod.Phys.,entem:2003_Phys.Rev.C,entem:2017_Phys.Rev.C,epelbaum:2015_Phys.Rev.Lett.}. Within this approach, the 3-nucleon interaction is only a small perturbation, arising at the third order of the perturbative scheme.

Although not as directly linked to the QCD parameters as the chiral limit, another critical point can be identified in the parameter space, corresponding to the unitary limit. In this case the separation of scales is provided by the large scattering lengths, resulting in a different (pionless) EFT \cite{vankolck:1999_Nucl.Phys.A,kaplan:1998_Phys.Lett.B,birse:1999_Phys.Lett.B,chen:1999_Nucl.Phys.A}. The two low-energy expansion schemes are different. In particular, for the reasons already explained, in the pionless EFT the 3-nucleon force is part of the LO description.

The question of the actual importance of the 3-nucleon force depends on which one of the two critical points can be considered as closer to the physical point. Furthermore, while the chiral regime of very small quark masses is outside of the Efimov window, ruled by the behaviour in the unitary limit, because the scattering lengths are natural in that limit \cite{epelbaum:2003_Nucl.Phys.A}, the unitary regime is met for values of the pion mass around 200~MeV \cite{epelbaum:2006_Eur.Phys.J.C,beane:2002_Nucl.Phys.A}  where the ChEFT should still apply. This means that the ChEFT treatment of the 3-nucleon force could have to be modified accordingly, by promoting it to the LO, as required by Efimov physics \cite{kievsky:2017_Phys.Rev.C}.
Indeed, although formally consistent, the ChEFT expansion scheme would fail in reproducing the universal correlations arising at the unitary limit, unless very high orders in the low-energy expansion are reached, so as to include the needed 3-nucleon force.

In order to study the impact of the explicit inclusion of the pion-range interactions on the sensitivity to the short-distance dynamics which was discussed previously, we make use
of the following lowest order Hamiltonian \cite{kievsky:2017_Phys.Rev.C}
\begin{equation}
H_{LO}=T+ \sum_{i<j}\left[V_{sr}(i,j)+V_\pi(i,j)+V_{EM}(i,j)\right] +\sum_{i<j<k} W(i,j,k)
\label{eq:hlo}
\end{equation}
where $T$ is the kinetic energy, $V_{sr}$ is the (regularized) short-range interaction introduced in
Eq.(\ref{eq:gauss3}), $V_{EM}$ is the electromagnetic interaction and $V_\pi$ is the OPEP
\begin{equation}
V_\pi(r)= {\bm \tau}_1\cdot{\bm \tau}_2
\left[ {\bm \sigma}_1\cdot{\bm \sigma}_2 Y_\beta(r)+ S_{12} T_\beta(r)\right] 
\end{equation}
with the regularized factors ($x=m_\pi r$)
\begin{eqnarray}
&Y_\beta(x)= {\displaystyle\frac{g_A^2 m_\pi^3}{12\pi F_\pi^2}
\,\frac{e^{-x}}{x}\left(1-e^{-r^2/\beta^2 }\right)} \\
&T_\beta(x)= \displaystyle{\frac{g_A^2 m_\pi^3}{12\pi
F_\pi^2}\,\frac{e^{-x}}{x}\left(1+\frac{3}{x}+\frac{3}{x^2}\right)\,\left(1-e^{-r^2/\beta^2}\right)^2\;\;} .
\end{eqnarray}
Here $m_\pi=138.03\;$MeV is the average pion mass, $g_A=1.29$ is the nucleon
axial coupling constant and $F_\pi=2f_\pi=184.80\;$MeV is the pion decay constant.
The regularization parameter $\beta$ is used to smoothly relate the chiral LO Hamiltonian 
to the pionless ($\beta\rightarrow\infty$) LO Hamiltonian.
Moreover, $H_{LO}$ includes a three-body term of the form
\begin{equation}
W(i,j,k)=W_0 e^{-r_{ij}^2/r_3^2} e^{-r_{ik}^2/r_3^2} \;\; .
\label{eq:w30}
\end{equation}
with $r_3$ the three-body range, $r_{ij}=|{\bm r}_i-{\bm r}_j|$, 
and the sum in Eq.(\ref{eq:hlo}) includes cyclic permutations of the three particles. 

\subsection{The excited $0^+$ state of $^4$He}
\begin{figure}
  \begin{center}
    \includegraphics[width=0.7\linewidth]{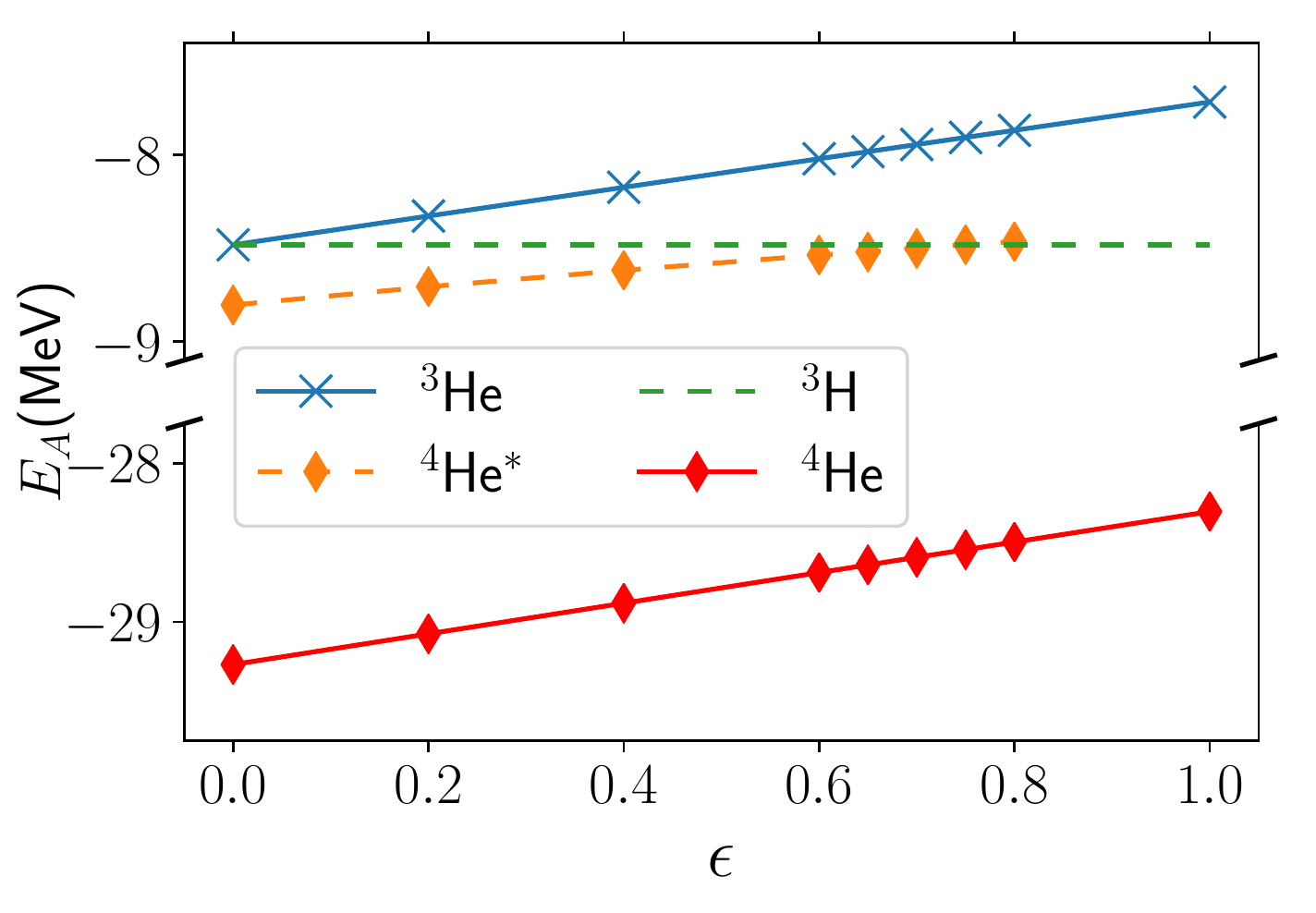}
  \end{center}
  \caption{Evolution of the $A=3,4$ energies as a function of a smooth
    switching-on of the Coulomb interaction via the multiplicative parameter
    $\epsilon$. The full Coulomb interaction corresponds to
    $\epsilon=1$.  The four-body excited state disappears at the critical value
    $\epsilon^*=0.754$. At $\epsilon=1$ the experimental energies of $^3$H, $^3$He and $^4$He
    are reproduced within a 1\% accuracy. Adapted from Ref.\cite{gattobigio:2019_Phys.Rev.C} with permission.}
  \label{fig:coulomb}
\end{figure}

As a first application of the pionless LO Hamiltonian we study the evolution of the
$^4$He excited state turning on adiabatically the Coulomb interaction
by considering $V_{EM}=\epsilon e^2/r$. 
The parameters of the two-body potential are fixed to reproduce the $np$ scattering
length and effective range in channels $S,T=0,1$ and $1,0$ whereas the
three-body term is fixed to describe the binding energy of $^3$H. Turning on smoothly the Coulomb interaction varying $\epsilon$
from 0 to 1, the two-body potential does not changes whereas the strength of the
three-body term is modified, maintaining its range fixed, to reproduce the
triton energy at each step. The results are shown in Fig.~\ref{fig:coulomb}
where, for $\epsilon=0$, we observe one $A=3$ state and the two $A=4$ states
~\cite{hammer:2007_Eur.Phys.J.A,gattobigio:2019_Phys.Rev.C,gattobigio:2012_Phys.Rev.A,deltuva:2013_Few-BodySyst,vonstecher:2009_NatPhys,
gattobigio:2013_Few-BodySystems}.
As the value of the Coulomb interaction grows to its
full value, $\epsilon=1$, the degeneracy between the $^3$H and $^3$He is
removed and the values of the ground- and excited-state energies of $^4$He
change. For $\epsilon\approx 0.75$ the $^4$He excited
state disappears onto the $^3$H+p threshold;
a polynomial fit gives the critical value  at $\epsilon^*=0.754$. At
$\epsilon=1$ the correct low-energy three- and four-nucleon spectrum is
recovered~\cite{gattobigio:2019_Phys.Rev.C}.

\subsection{The saturation point of nuclear matter}
The application of the $H_{LO}$, given in Eq.(\ref{eq:hlo}) to the case of $A=3,4$ and nuclear matter is 
extensively discussed in
Refs.~\cite{kievsky:2017_Phys.Rev.C,kievsky:2018_Phys.Rev.Lett.}. The energy per nucleon of nuclear matter is calculated 
using the Brueckner--Bethe--Goldstone (BBG) quantum many-body theory in the
Brueckner--Hartree--Fock (BHF) approximation (see e.g. \cite{bombaci:2018_Astron.Astrophys., logoteta:2016_Phys.Rev.C,logoteta:2016_Phys.Lett.B} and references therein).
In the calculations the three-nucleon force has been reduced to an effective,
density dependent two-body force, by averaging over the coordinates of the third
nucleon~\cite{logoteta:2016_Phys.Rev.C}.

\begin{figure}
\includegraphics[scale=0.5]{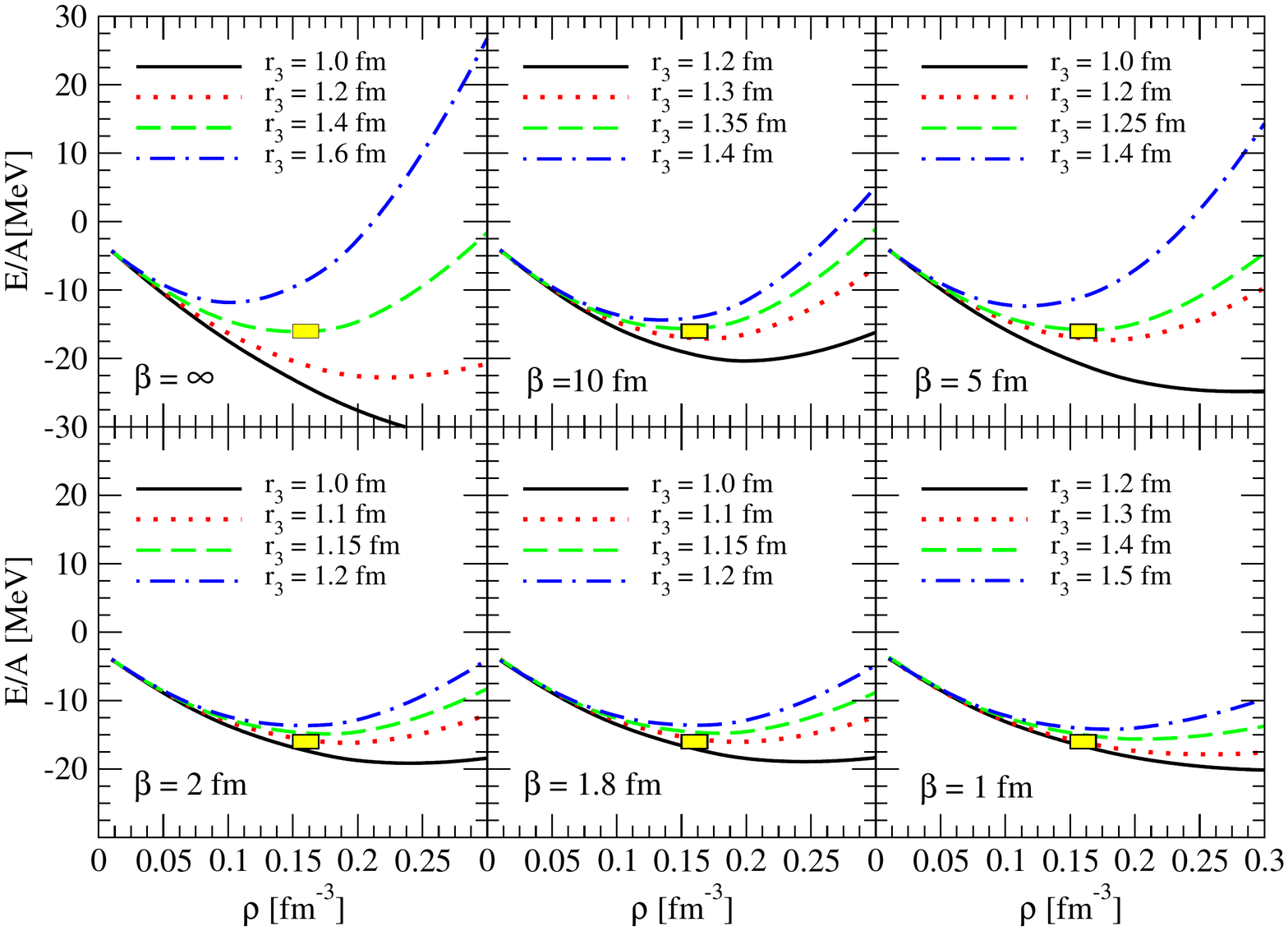}
\caption{Energy per particle of symmetric nuclear matter $E/A$ as a function of
the nucleonic
density $\rho$ for several combinations of the LO two- and three-body
interactions. The yellow box denoted the empirical saturation point. Adapted from Ref.\cite{kievsky:2018_Phys.Rev.Lett.} with permission.}
\label{fig:nmatter}
\end{figure}

The energy per particle $E/A$ of symmetric nuclear matter (SNM) is shown in
Fig.\ \ref{fig:nmatter} for various parametrizations of the two- and three-body forces. In each panel,
for a fixed value of the OPEP regulator $\beta$ of the two-body force, the saturation
curve (i.e. $E/A$ as a function of the nucleonic density $\rho$) of SNM is shown using
four different values of the three-nucleon force range $r_3$, determined to
describe the $^3$H binding energy.
The empirical saturation point of SNM ($\rho_{0} = 0.16 \pm  0.01~{\rm
fm}^{-3}$, $E/A|_{\rho_0} = -16.0 \pm 1.0~{\rm MeV}$) is denoted by a yellow box in each
panel of Fig.\ \ref{fig:nmatter}. Interestingly, the saturation point is well
described for values of $r_3$ compatible with a correct description of $^4$He,
with the best description obtained in the pionless case,
$\beta\rightarrow\infty$. This is another example of strict correlations, in
this case for the nuclear system,  between the low-energy few-body properties and the many-body system, which are induced by the physics of the universal window.

\begin{summary}[SUMMARY POINTS]
\begin{enumerate}
    \item The Gaussian characterization of the univeral window discussed in this article  is based on the simplest description of the one level two-body $S$-matrix\cite{babenko:2000_Phys.At.Nucl.,newton:1982_}
\begin{equation}
    S(k)=\frac{k+i/a_B}{k-i/a_B}\frac{k+i/r_B}{k-i/r_B},
\end{equation}
equivalent to the ERE of Eq.(\ref{eq:ERE}). Inside the window, defined by condition $r_B \ll a_B$, we have selected a Gaussian potential to reproduce this behavior in the two-body sector and used it to extend the description to larger systems. 

\item Two-body systems manifest a CSI, depicted in Fig.~\ref{fig:fig1}, which is broken to a DSI in three-body systems. The adopted procedure allows to address the impact of finite range corrections on three-body levels. Interestingly,  only the first two levels are affected in a significant way, with the higher levels tending rapidly to the zero range limit. The position of a physical system on the lowest level is controlled by a three-body datum. Then its spectrum and correlations with the low-energy scattering states are
completely determined. Examples have been shown for two very different systems, the helium trimer and the three-nucleon system. Furthermore, the study has been extended to larger systems showing how they are still constrained by the DSI.

\item We have highlighted a number of properties, for systems belonging to the universal window, that can be understood as consequences of their position inside the window. The present analysis suggests that the EFT describing nuclear interactions should incorporate a three-nucleon term at LO, independently if the pions are integrated out or not.
This important consequence is based on the extreme sensitivity of the three-nucleon system to the cutoff effects at LO.
The inclusion of a three-nucleon force at LO in the nuclear hamiltonian will have significant consequences in the description of nuclei using precise interactions derived from ChEFT
\cite{entem:2017_Phys.Rev.C,epelbaum:2015_Phys.Rev.Lett.,kievsky:2017_Phys.Rev.C,girlanda:2019_Phys.Rev.C,girlanda:2020_Phys.Rev.C}. 

\item As the Gaussian characterization is used to describe systems with larger number of particles, system-specific non-universal behavior starts to emerge~\cite{kievsky:2020_Phys.Rev.A,kievsky:2017_Phys.Rev.A}. 
Indeed, the position of a system inside the universal window determines the two-body Gaussian potential from the values of $a_B$ and $a$ whereas the three-body binding energy determines the strength of the three-body potential. The use of this two- plus three-body potential to describe heavier systems introduces a dependence on a short range scale which is a non universal effect. This effect can be incorporated tuning the range of the three-body interaction, a parameter that can be used to improve the convergence of the EFT expansion, or by including higher orders.

\item Efimov physics has substantial implications for the dynamical description of systems located inside the universal window. These systems are strongly constrained by an (approximate) scale invariance. A thorough analysis of its consequences in the many-body sector is an important task which is at present intensively pursued.

\end{enumerate}
\end{summary}
\section*{DISCLOSURE STATEMENT}
The authors are not aware of any affiliations, memberships, funding, or financial holdings that might be perceived as affecting the objectivity of this review

\bibliographystyle{ar-style5.bst}

\end{document}